\documentclass[useAMS,usenatbib]{mn2e}
\usepackage[T1]{fontenc}
\usepackage{graphicx}
\usepackage{amsmath,times,amssymb,calligra,amsbsy}
\DeclareMathAlphabet{\mathcalligra}{T1}{calligra}{m}{n}
\DeclareFontShape{T1}{calligra}{m}{n}{<->s*[2.2]callig15}{}
\usepackage{macros}
\usepackage[usenames,dvipsnames,svgnames,table]{xcolor}
\usepackage{hyperref}
\definecolor{darkblue}{rgb}{0.0,0.0,0.5}
\definecolor{darkred}{rgb}{0.5,0.0,0.0}
\hypersetup{colorlinks,breaklinks,
            linkcolor=darkblue,urlcolor=darkblue,
            anchorcolor=darkblue,citecolor=darkblue}

\newcommand{\appropto}{\mathrel{\vcenter{
  \offinterlineskip\halign{\hfil$##$\cr
    \propto\cr\noalign{\kern2pt}\sim\cr\noalign{\kern-2pt}}}}}
\DeclareSymbolFont{cmletters}{OML}{cmm}{m}{it}

\newcommand\bb[1]{\mbox{\boldmath{$#1$}}}
\newcommand\grad{\bb{\nabla}}
\newcommand\bcdot{\bb{\cdot}}

\newcommand{\msb}[1]{\bb{\mathsf{#1}}}

\title[Three-Dimensional Disk-Satellite Interaction]{Three-Dimensional Disk-Satellite Interaction: Torques, Migration, and Observational Signatures}

\author[Lev Arzamasskiy, Zhaohuan Zhu, James M. Stone]{Lev Arzamasskiy$^{1}$\thanks{E-mail:
leva@astro.princeton.edu}, Zhaohuan Zhu$^{2}$, James M. Stone$^{1}$\\
$^{1}$Department of Astrophysical Sciences, 
Princeton University, Princeton, NJ 08540\\
$^{2}$Department of Physics and Astronomy, UNLV, Las Vegas, NV 89154
}
\begin{document}

\date{Accepted. Received; in original form}
\pagerange{\pageref{firstpage}--\pageref{lastpage}} \pubyear{2017}

\maketitle

\label{firstpage}

\begin{abstract}
The interaction of a satellite with a gaseous disk results in the excitation of spiral density waves which remove angular momentum from the orbit.  In addition, if the orbit is not coplanar with the disk, three-dimensional effects will excite bending and eccentricity waves. We perform three-dimensional hydrodynamic simulations to study nonlinear disk-satellite interaction in inviscid protoplanetary disks for a variety of orbital inclinations from $0^\circ$ to $180^\circ$. It is well known that three-dimensional effects are important even for zero inclination. In this work we (1) show that for planets with small inclinations (as in the Solar system), effects such as the total torque and migration rate strongly depend on the inclination and are significantly different  (about 2.5 times smaller) from the two-dimensional case, (2) give formulae for the migration rate, inclination damping, and precession rate of planets with different inclination angles in disk with different scale heights, and (3) present the observational signatures of a planet on an inclined orbit with respect to the protoplanetary disk. 
For misaligned planets we find good agreement with linear theory in the limit of small inclinations, and with dynamical friction estimates for intermediate inclinations. We find that in the latter case, the dynamical friction force is not parallel to the  relative planetary velocity. Overall, the derived formulae will be important for studying exoplanets with obliquity.

\end{abstract}

\begin{keywords}
planets and satellites: formation --- protoplanetary disks ---
stars: planetary systems
\end{keywords}
\section{Introduction}  
\label{sect:intro}

The interaction between a gaseous disk and an embedded satellite can have a major effect on the dynamics and structure of the disk. The study of such interaction was pioneered by \citet{1979ApJ...233..857G,GT80}, who found that density waves are excited due to the gravitational forces between the secondary component and the disk. These forces lead to the exchange of energy and angular momentum between the disk and perturber, and thus to the change of the orbital elements of the latter.  The linear, two-dimensional calculations of \citet{GT80} have been extended in a number of ways, e.g. by considering the asymmetry between inner and outer disk \citep{1986Icar...67..164W}, and to the case of fully three-dimensional (3D) modes in disks with vertical structure \citep{2002ApJ...565.1257T, 1993ApJ...419..166A}.

The non-zero inclinations of planets in the Solar system as well as a
growing number of observations of misaligned extrasolar planets motivates further extensions to the theory to take into account a non-zero inclination of the secondary component. It was found that misaligned planets warp the disk and excite bending waves \citep{1995ApJ...438..841P,1996A&A...307...21M,1997A&A...324..829D,2000ApJ...538..326L,2010A&A...511A..77F}. Bending waves also participate in the exchange of angular momentum between the disk and perturber, and make the evolution of the satellite's orbital elements more complex.

\citet{2004ApJ...602..388T} described the interaction between a disk and an inclined planet in the linear regime and found an exponential damping of inclination on a timescale of $\sim 10^3$ years for an Earth-sized planet at $1$~AU in the minimum mass solar nebula. However, to be valid the theory formally requires the inclination of the planet $i_{\rm p}$ to be much smaller than the aspect ratio of the disk, that is $i_{\rm p} \ll H/r$, where $H$ is the thermal scale height in the disk.  Numerical studies \citep{2007A&A...473..329C,2011A&A...530A..41B} have confirmed the exponential damping and extended the applicability range to $i_{\rm p} \la 2 H/r$. For even larger inclinations, these studies find the damping rate ${\rm d} i_{\rm p}/{\rm d}t \propto i_{\rm p}^{-2}$.  However, the analytic calculations of \citet{2012MNRAS.422.3611R}, which include dynamical friction as the dominant mechanism of angular momentum transport, predict a different scaling ${\rm d} i_{\rm p}/{\rm d}t \appropto i_{\rm p}^{-3}$.

The goals of this paper are (1) to verify the results of linear theory \citet{2002ApJ...565.1257T,2004ApJ...602..388T} for the case of planets with small inclinations, (2) to investigate the evolution of planets with very large inclinations and to resolve the discrepancy between current numerical and analytic models, and (3) to predict the observational signatures of misaligned planets in gaseous protoplanetary disks. 

Our study uses high resolution 3D numerical simulations of planet-disk interaction for perturbers with small masses $M_{\rm p}/M_\star = 10^{-4}$ (where $M_{\rm p}$ and ${M_\star}$ are the masses of the planet and central star respectively).  Thus, the amplitude of the perturbations are well within the linear regime.  The disk is assumed to be globally isothermal (consistent with the assumptions used in linear theory), while the inclination of the perturber is varied over a wide range $0.25 H/r \leq i_{\rm p} \leq \pi$, that is we study the case of satellites on both perpendicular as well as retrograde orbits. To predict the observational signatures, we carry out simulations with a locally isothermal disk and Jupiter-mass planets, and compare the near-IR scattered light images for both coplanar and misaligned planets.

The paper is organized as follows. We start by discussing the setup of our numerical calculations in Section \ref{sect:setup}. Section \ref{sect:test} reports the result from a sensitive test of the hydrodynamics code used for this study; the propagation of bending waves in the linear regime.  We then present results from fully 3D calculations of planets on coplanar orbits in Section \ref{sect:coplanar}, including analysis of the torque density compared to previous analytical results. Section \ref{sect:inclined} describes the evolution of the orbital elements for inclined planets, the main result of this paper. The response of the disk to the planetary potential as well as the observational signatures of inclined planets in scattered light images are discussed in Section \ref{sect:obs}. Finally, we summarize our results and conclude in Section \ref{sect:conclusion}.

\section{Simulation setup}
\label{sect:setup}

\subsection{Disk structure}
\label{sect:disk}
To study disk-planet interaction in 3D, we run a set of hydrodynamical
simulations in spherical polar coordinates $(r,\theta,\varphi)$ using Athena++ (Stone et al. 2017, in preparation). To make a direct comparison
with analytical results, we consider a globally isothermal disk. This is 
unlikely to be the case in real protoplanetary disks, and we use the locally isothermal approximation
instead when constructing scattered light images in Sect.~\ref{sect:obs}.  Thus, we solve
\begin{align}
\frac{\partial \rho}{\partial t} + \grad \bcdot (\rho \bb{v}) &= 0,\\
\frac{\partial \rho \bb{v}}{\partial t} + \grad \bcdot (\rho \bb{v} \bb{v} + \msb{P}) + \rho \grad \phi_{\rm gr} &= 0,\\
\frac{\partial E}{\partial t} + \grad \bcdot [(E+P) \bb{v}] + \mathcal{Q} &= 0,\label{eq:energy_eq}
\end{align}
where $\rho$ is the mass density of the fluid, $\boldsymbol{v}$ is fluid velocity, $\msb{P}$ is a diagonal pressure tensor with components equal to $P$, and $E$ is the total energy of the fluid:
\begin{equation}
    E = \frac{P}{\gamma - 1} + \frac{1}{2}\rho v^2 + \rho \phi_{\rm gr}.
\end{equation}
Gravitational potential $\phi_{\rm gr} = \phi_\star + \delta \phi$ consists of the potential of the central object $\phi_\star = -GM_\star/r$ and perturbation potential caused by the planet (which we discuss in Section \ref{sect:planet}).
For globally isothermal simulations, equation (\ref{eq:energy_eq}) is not solved and pressure is assumed to be $P = c_{\rm s}^2 \rho$ with constant sound speed $c_s$. For locally isothermal simulations, we use $\gamma = 5/3$ and $c_{\rm s} = c_{\rm s}(r)$. In this case, we solve energy equation (\ref{eq:energy_eq}), but to make the disk locally isothermal, we implement a cooling source term $\mathcal{Q}$.  The detailed setup of our cooling prescription is given in \citet{2015ApJ...813...88Z}. We use a finite cooling time $T_{\rm cool} = 1/\Omega_{\rm p}$. Note that we do not include any self-gravity or explicit viscosity in
our calculations.  Thus, our results are applicable to low-mass protoplanetary disks with very low accretion 
rates (and therefore effective viscosity).

In order to obtain initial conditions for our simulations we use the equilibrium disk model
described by \cite{2013MNRAS.435.2610N}. This model has two parameters $q$ and $p$ which fix the initial mid-plane density and temperature profiles:
\begin{align}
T^{\rm in}(r,\theta) &= T_0 (r\sin\theta/r_0)^q,\label{eq:temp_in}\\
\rho_{\rm mid}^{\rm in}(r,\theta) &= \rho_0 (r\sin\theta/r_0)^p,\label{eq:middens_in}
\end{align}
where $T_0$, $\rho_0$, and $r_0$ are the characteristic values of temperature, density, and size of the disk.

The power-law profiles (\ref{eq:temp_in})-(\ref{eq:middens_in}) imply the following exact analytic equilibrium solution of the Euler equations for a non self-gravitating disk around a central star:
\begin{align}
\rho^{\rm in} &= \rho_{\rm mid}^{\rm in}(r,\theta) \exp\left[\frac{GM_\star}{r c_{\rm s}^2}\left(1 - \frac{1}{\sin\theta} \right) \right],\label{eq:dens_in}\\
\Omega^2 &= \Omega_{\rm K}^2\left[(p+q)\left(\frac{c_{\rm s}/\Omega_{\rm K}}{r\sin\theta}\right)^2 + 1+q(1-\sin\theta) \right],\label{eq:omega_in}
\end{align}
where $c_{\rm s}(r,\theta)$ is the sound speed, $c_{\rm s}^2/c_{{\rm s},0}^2 = T/T_0$, and $M_\star$ is the mass of central star. The scale-height of the disk $H$ is
\begin{equation}
H \equiv c_{\rm s}/\Omega_{\rm K}|_{\theta = \pi/2} = H_0 (r/r_0)^{(q+3)/2}.
\end{equation}
The Keplerian angular velocity $\Omega_{\rm K}$ in this model is
\begin{equation}
\Omega_{\rm K}^2 = GM_\star/(r\sin\theta)^3, 
\end{equation}
corresponding to the initial $\varphi$-component of velocity:
\begin{equation}
v_{\varphi}^{\rm in} = \sqrt{GM_\star/(r\sin\theta)}.
\end{equation}

Most of our simulations assume a {\it globally} isothermal disk, for which $q=0$ and the aspect ratio
\begin{equation}
h\equiv H/r \propto r^{1/2}. \quad {\rm (globally~isothermal)}
\end{equation}
In realistic protoplanetary disks, conservation of radiation flux implies $T^4 r^2 = {\rm const}$ in equilibrium (see also \citealt{1997ApJ...490..368C} for other possibilities in more realistic disk models). This case corresponds to the {\it locally} isothermal disk, for which $q = -1/2$ and
\begin{equation}
h \equiv H/r \propto r^{1/4}. \quad {\rm (locally~isothermal)}
\end{equation}

In what follows, we define the surface density of the disk as the following integral:
\begin{equation}
{\Sigma}(r) \equiv \frac{1}{2\pi} \int\limits_0^{2\pi}\int\limits_{-\infty}^{\infty}\rho {\rm d}z{\rm d}\varphi,
\end{equation}
where $z = r \cos \theta$.

The initial density profile (\ref{eq:dens_in}) has a large and narrow peak at $\theta = 0$ for small $h  \ll 1$:
\begin{align}
\rho^{\rm in}(r,\theta) &\approx \rho_0 (r/r_0)^p \exp\left(-\frac{(\theta-\pi/2)^2}{2 h^2} \right),~{\rm and}\\
\Sigma(r) &\approx \sqrt{2\pi}\rho_0 H_0 (r/r_0)^{p+(q+3)/2} =\\
&= \Sigma_0 (r/r_0)^{p+(q+3)/2}.\nonumber
\end{align}

We set the average surface density to decrease as $ \Sigma \propto r^{-1/2}$, so for a globally isothermal disk we use $ p = -2$, while for a locally isothermal disk with $q = -0.5$, we use $p = -1.75$.

\subsection{Planetary potential}
\label{sect:planet}

To study the effect of the planet on the disk we introduce a perturbation potential
\begin{equation}
\delta\phi = \delta\phi_{\rm p}^{(6)} + \frac{M_{\rm p}}{M_\star}\Omega_p^2(\boldsymbol{r}_p\cdot\boldsymbol{r}),
\end{equation}
where the second term is the indirect potential caused by the motion of the primary star, $M_{\rm p}$ and $M_\star$ are the masses of the planet and the central star, and
\begin{equation}
\delta\phi_{\rm p}^{(6)} = - GM_{\rm p}\frac{(\boldsymbol{r} - \boldsymbol{r}_{\rm p})^4 + 2.5 (\boldsymbol{r}-\boldsymbol{r}_{\rm p})^2 R_{\rm sm}^2 + 1.875 R_{\rm sm}^4}{[(\boldsymbol{r}-\boldsymbol{r}_{\rm p})^2 + R_{\rm sm}^2]^{5/2}}
\end{equation}
is the sixth-order accurate smoothed gravitational potential of the planet (the fractional error is $\mathcal{O}(R_{\rm sm}^6/|\boldsymbol{r}-\boldsymbol{r}_{\rm p}|^6)$ as $R_{\rm sm}/|\boldsymbol{r}-\boldsymbol{r}_{\rm p}| \rightarrow 0$, see \citet{2011ApJ...741...56D} for the discussion of accuracy of different smoothing prescriptions).

We evolve the system for a time which is considerably smaller than the migration timescale.  Thus, we fix the planetary orbit $\boldsymbol{r}_{\rm p}(t) = \rm{const}$. For a planet with arbitrary inclination $i_{\rm p}$, its orbit in Cartesian coordinates is
\begin{equation}
\boldsymbol{r}_{\rm p} = (r_{\rm p} \cos i_{\rm p}\cos\Omega_{\rm p} t, r_{\rm p}\sin\Omega_{\rm p} t, r_{\rm p} \sin i_{\rm p} \cos\Omega_{\rm p} t).
\end{equation}

To obtain torques acting on the planet, we run our simulations for $10-15$ orbital periods. We found that the torque usually saturates after $2-3$ orbits. After saturation we average the torque over $5-10$ orbits.

\section{Code test: propagation of a bending wave}
\label{sect:test}

\begin{table}
\caption{Parameters for bending wave propagation test. For all runs we use $q = 0$, $p = -1.5$, the azimuthal domain is $[0,2\pi]$, and the maximum warp amplitude is located at $r_{\rm mid}/r_0 = 5$. The table lists aspect ratio of the disk at inner boundary $H_0/r_0$, the amplitude of the warp $\delta_{\rm max}$, size of the simulation domain, and resolution for each run.\label{tabl:warp}}
\begin{tabular}{cccc}
 $H_0/r_0$ & $\delta_{\rm max}$ & Domain $[r/r_0]\times[\theta]$ &  $N_r\times N_\theta\times N_\varphi$\\
\hline
\hline
  $0.035 = 2^\circ$ & $0.017=1^\circ$  & $[1,9]\times[70^\circ, 110^\circ]$ & $384\times 128 \times 1024$\\
 $0.035= 2^\circ$ & $0.035= 2^\circ$ & $[1,9]\times[70^\circ, 110^\circ]$ & $384\times 128 \times 1024$\\
 $0.070= 4^\circ$ & $0.035= 2^\circ$ & $[1,9]\times[50^\circ, 130^\circ]$ & $384\times 256 \times 1024$
\end{tabular}
\end{table}

\begin{figure*}
\centering
\includegraphics[width=\columnwidth]{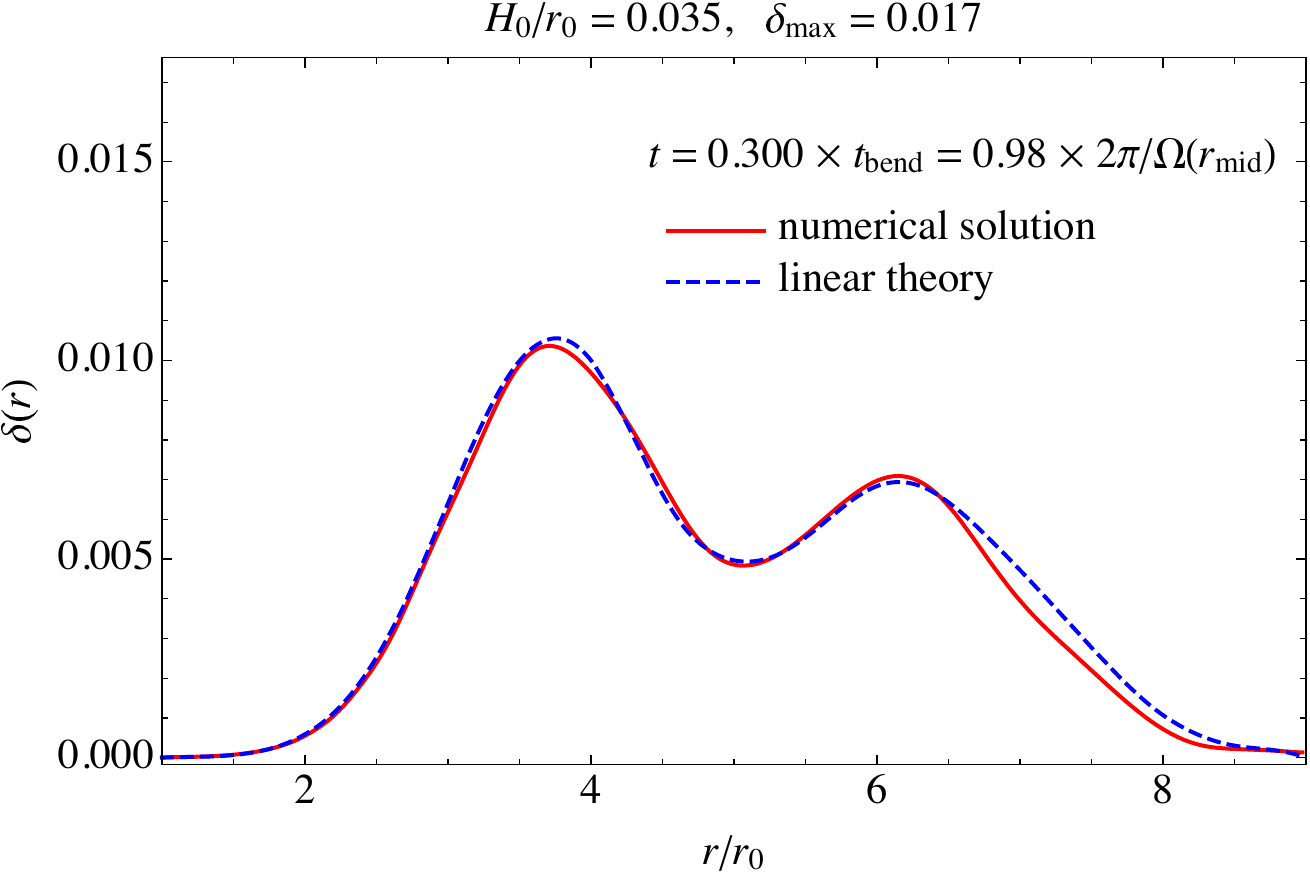}~~~~~\includegraphics[width=\columnwidth]{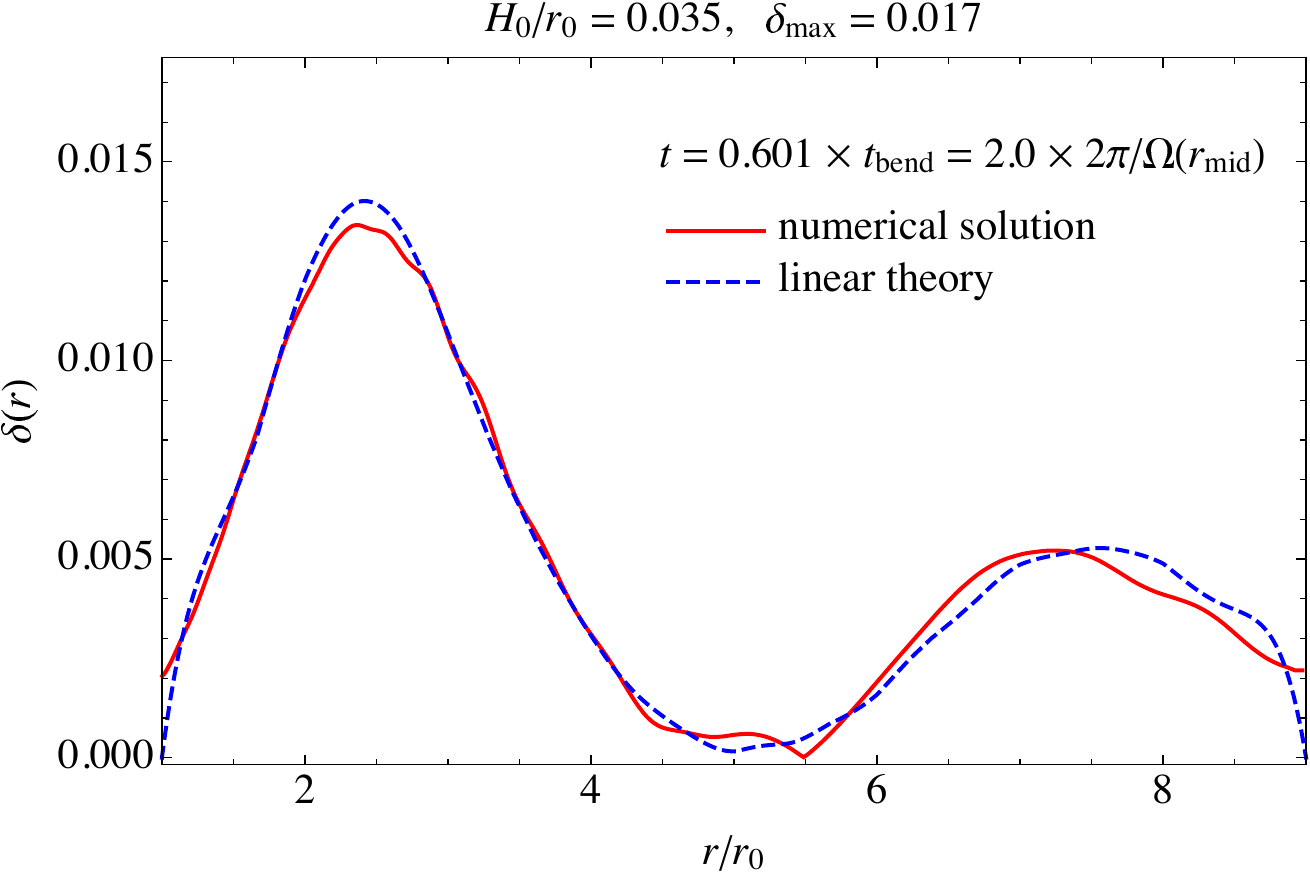}
\caption{Bending wave propagation for a thin disk ($H_0/r_0  = 0.035 = 2^\circ$) with initial warp amplitude of $1^\circ$. The numerical solution is in very good agreement with estimates from analytic linear theory.}
\label{fig:warp_lin}
\end{figure*}

For the general case of disk-planet interaction in 3D, a planet on an inclined orbit induces bending waves in the disk. In this Section, we test our code against linear theory to determine how well it is able to treat bending wave propagation. We perform the test described by \citet{2010A&A...511A..77F} which models the propagation of bending waves in an initially warped disk with no planet, and compare our solution with estimates from linear theory \citep{1995ApJ...438..841P, 1996A&A...307...21M,1997A&A...324..829D,2000ApJ...538..326L}.

In what follows we describe the equations used to study bending waves in cylindrical coordinates. Our numerical simulations used for comparison are still performed in spherical coordinates.

The propagation of bending waves in the linear regime can be described in cylindrical coordinates $(R,~\phi,~z)$ by the following set of partial differential equations \citep{1999MNRAS.309..929N}:
\begin{align}
&\partial_t q_R + i \Omega q_\phi = i \partial_R(R \Omega^2 \delta),\label{eq:pn1999begin}\\
&\partial_t q_\phi + i \Omega q_\phi +q_R\partial_R(R^2 \Omega)/R = -\Omega^2 \delta,\\
&\partial_t v_z + i\Omega v_z = i z \Omega^2 \delta,\\
&\mathcal{F}R\Omega^2 (\partial_t\delta + i \Omega\delta) = - i\partial_R(R\mu q_R)/R +\mu q_\phi/R
 + i\Sigma v_z. \label{eq:pn1999end}
\end{align}
Here $v_R$, $v_\phi$ and $v_z$ are perturbations in each of the velocity components of the disk, which is initially in equilibrium.  Here $q_R = v_R/z$, $q_\phi = v_\phi/z$, and the quantity $\delta$ describes the local inclination angle which is directly related to the density perturbation $\delta\rho$:
\begin{equation}
\frac{\delta\rho}{\rho} c_{\rm s}^2 = - i R z \Omega^2 \delta.
\end{equation}
Finally, the functions $\mathcal{F}$ and $\mu$ are given by the following integrals:
\begin{equation}
\mathcal{F} = \int\limits_{-\infty}^{+\infty} \frac{\rho z^2}{c_{\rm s}^2}{\rm d}z,~{\rm and}~ \mu = \int\limits_{-\infty}^{\infty} \rho z^2 {\rm d}z.
\end{equation}

It is important to understand the limits in which the system (\ref{eq:pn1999begin})--(\ref{eq:pn1999end}) applies. To derive this system, it is necessary to assume a very thin disk, so $\Omega$ and $c_{\rm s}$ do not depend on $z$. Next, the perturbation amplitude must be small. This implies the following restrictive relation:
\begin{equation}
\delta \ll H/r \ll 1.
\end{equation}
This relation is hard to satisfy in numerical simulations, and it is often the main source of disagreement between analytic and numerical solutions. Finally, the derivation of the system (\ref{eq:pn1999begin})-(\ref{eq:pn1999end}) assumes a barotropic equation of state $p = p(\rho)$, which is true for globally (but not locally) isothermal disks.

We solve the system (\ref{eq:pn1999begin})-(\ref{eq:pn1999end}) with reflecting boundary conditions
\begin{equation}
\textbf{v},~\delta = 0, ~{\rm at}~R=R_{\rm in},~R_{\rm out};
\end{equation}
and the following initial conditions \citep{2010A&A...511A..77F}:
\begin{align}
v_R &= - z \Omega \delta_0(R)\label{eq:warp_init_begin},\\
v_\phi &= -iz\delta_0(R) {\rm d}(R \Omega)/{\rm d}R,\\
v_z &= R\Omega \delta_0(R),\\
\delta &= \delta_0(R)\label{eq:warp_init_end},
\end{align}
where $\delta_0(R)$ is the initial inclination angle. If $\delta_0 = {\rm const}$, the disk has rigid inclination and has no time-dependence. But for $\delta_0 \ne {\rm const}$, bending waves start to propagate.

In what follows we use 
\begin{equation}
\delta_0(R) = \delta_{\rm max} \exp\left[-(R - R_{\rm mid})^2\right],
\end{equation}
where $\delta_{\rm max}$ is the warp amplitude, and $R_{\rm mid}$ is the location of the warp center.  To simulate a warped disk, we slightly modify the initial equilibrium to account for the varying disk inclination:
\begin{align}
\rho(r,\theta) &\rightarrow \rho(r,\theta - \theta_{\rm mid}),\\
\Omega(r,\theta) &\rightarrow \Omega(r,\theta - \theta_{\rm mid}),
\end{align}
where $\theta_{\rm mid} = \theta_{\rm mid}(r ,\theta)$ can be determined from the following relation \citep{2010A&A...511A..77F}:
\begin{align}
\sin^2(\theta - \theta_{\rm mid}) &= \cos^2\varphi\sin^2\delta_0\sin^2\theta + \sin^2\delta_0\cos^2\theta-\nonumber\\
&- 2 \cos\delta_0\sin\delta_0\sin\theta\cos\theta\sin\varphi.
\end{align}
If $\delta_0 = 0$, then $\theta_{\rm mid} = 0$. 

Each component of the velocity in the disk is given by
\begin{align}
 v_r^{\rm in} &= 0, \\
 v_\theta^{\rm in} &= \Omega r \sin\delta_0 \cos\varphi,\\
 v_\varphi^{\rm in} &= \Omega r (\cos\delta_0\sin\theta - \sin\varphi\sin\delta_0\cos\theta).
\end{align}
These expressions correspond to an initial warp $\delta_0(r)$ in the $\hat x$ direction of a Cartesian coordinate system.

\begin{figure*}
\centering
\includegraphics[width=\columnwidth]{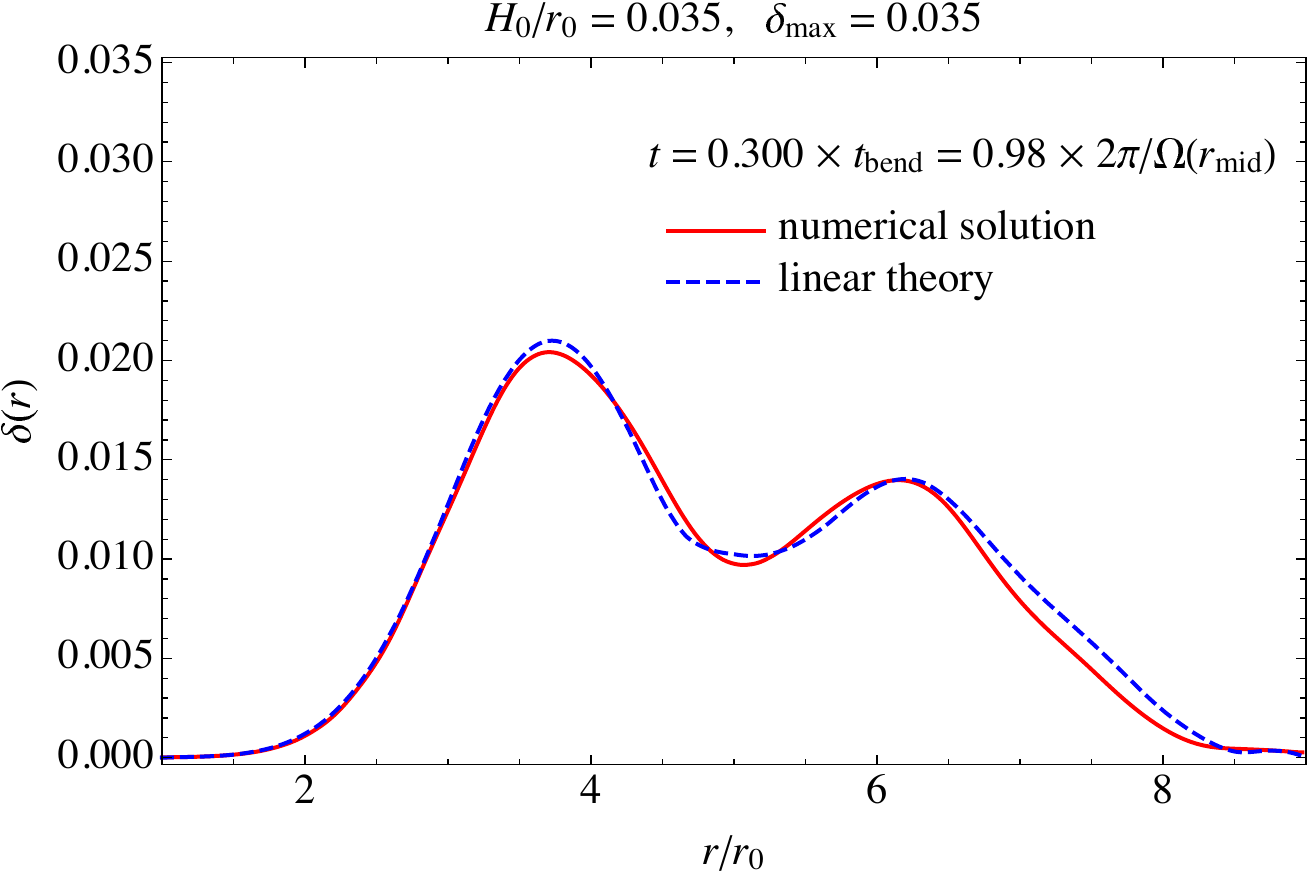}~~~~~\includegraphics[width=\columnwidth]{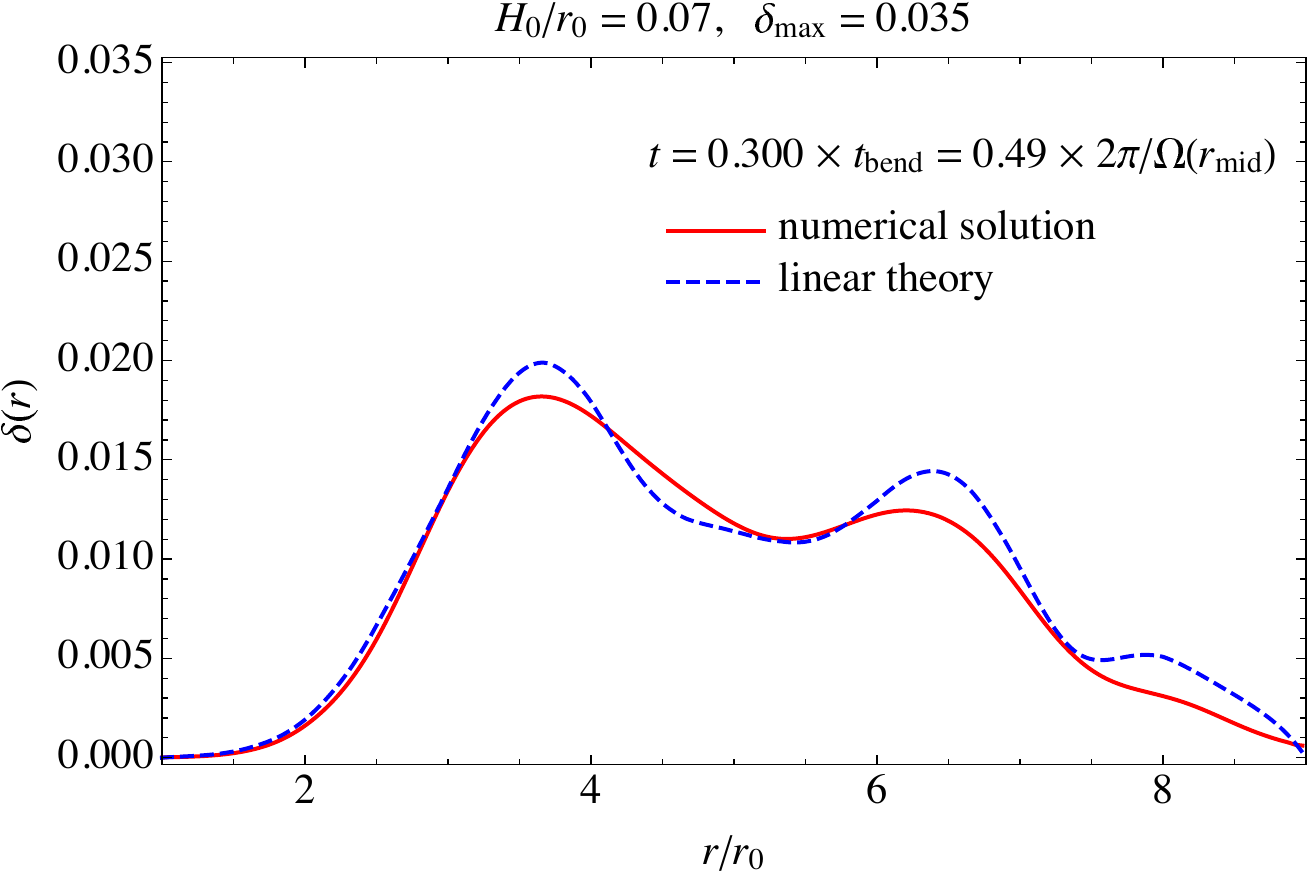}
\caption{ \textbf{[Left panel]:} Simulation results for a thin disk ($H_0/r_0  = 0.035 = 2^\circ$) with a large warp amplitude of $2^\circ$. The shape of the curve is very similar to the case with a smaller perturbation, thus the non-linearity caused by the warp is negligible. \textbf{[Right panel]:} Propagation of a warp in a thicker disk ($H_0/r_0  = 0.07 = 4^\circ$) with the same warp amplitude of $2^\circ$.  Now there are substantial differences in the profile of the inclination angle compared to the predictions of linear theory.  Thus, the non-linearity caused by the thickness of the disk is strongly affecting the solution.}
\label{fig:warp_nonlin}
\end{figure*}

To test our code, we study the propagation of a linear bending wave with $H_0/r_0 = 0.035$ (corresponding to $2^\circ$), $\delta_{\rm max} = 0.5 H_0/r_0$, and $r_{\rm mid} = 5 r_0$ (note that at the location of the center of the warp, $H/r = \sqrt{5} H_0/r_0$, so the warp amplitude is approximately 1/4 of the disk aspect ratio). The simulation setup used for this test is similar to Section \ref{sect:setup}, but without planetary potential, and with warped disk midplane. We use an isothermal equation of state ($q=0$) so that it is barotropic to enable comparison with the solution of equations (\ref{eq:pn1999begin}) -- (\ref{eq:pn1999end}). Following \citet{2010A&A...511A..77F}, we set $p=-1.5$ (even though their disk was locally isothermal, and the equilibrium solution was different). Our $\theta$-domain spans 10 aspect ratios at the inner boundary in each direction (note, that it corresponds to 10/3 aspect ratios at the outer boundary). We implement outflow boundary condition in $\theta$-direction. We set the warp midpoint to be $r_{\rm mid} =5r_0$, so the propagation time for bending waves is approximately the same in both directions (linear theory predicts the speed of bending waves to be $v_{\rm bend} = c_{\rm s}/2 = \rm const$ for isothermal disks). Please see Table \ref{tabl:warp} for more details of this simulation.

When studying the properties of the disk, it is useful to define the inclination and precession angles. We define the former as the angle between the total angular momentum of a shell with given radius $\boldsymbol{L}(r)$ and the unperturbed angular momentum of that shell (which in our case coincides with the $\hat z$-axis):
\begin{equation}
\label{eq:delta(r)}
\cos\delta(r)\equiv \frac{\boldsymbol{L}(r)\cdot\boldsymbol{e}_z}{|\boldsymbol{L}(r)|},
\end{equation}
and the latter as \citep{2010A&A...511A..77F}
\begin{equation}
\cos\beta(r) \equiv \frac{(\boldsymbol{L}(r)\times\boldsymbol{e}_z)\cdot \boldsymbol{e}_x}{|\boldsymbol{L}(r)\times \boldsymbol{e}_z|}.
\end{equation}

Figure~\ref{fig:warp_lin} illustrates the results of our code test.  The inclination angle of the disk at times approximately equal to $0.3 t_{\rm bend}$ and $0.6 t_{\rm bend}$ are shown, where $t_{\rm bend}$ is equal to the expected bending wave propagation time:
\begin{equation}
t_{\rm bend} = \frac{r_{\rm mid} - r_0}{c_{\rm s}/2}.
\end{equation}
We do not continue our solution beyond $0.6 t_{\rm bend}$ due to reflections from boundaries in the analytic solution. It is obvious from the figure that the numerical solution is in very good agreement with the analytic estimates, implying that Athena++ can reproduce the linear behavior not only of density waves for coplanar orbits as was demonstrated by \citet{2015ApJ...813...88Z,2015ApJ...809L...5D}, but also of bending waves excited by inclined orbits.

\begin{table}
\small
\caption{Simulation set up\label{tabl:nonlinear}}
\begin{tabular}{ll}
Parameters & Values\\
\hline
\hline
Domain & $[0.4, 2.0]\times [\pi/2-0.3,\pi/2+0.3]\times [0,2\pi]$\\
$N_r\times N_\theta \times N_\phi$ & $512\times192\times1920$\\
Mass of the planet & $M_{\rm p} = 10^{-4} M_\star = 0.1 M_{\rm th}$\\
Smoothing length & $R_{\rm sm} = 0.025 r_{\rm p} = 0.25 H_{\rm p} \approx 0.78 R_{\rm Hill}$\\
Resolution & $20/H_{\rm in} = 32 /H_{\rm p} = 10/R_{\rm Hill} = 8 /R_{\rm sm}$
 \end{tabular}
\end{table}

\subsection{Deviations from linear theory}
\label{sect:nonlin}

The test described in the previous section also allows one to investigate the limits in which linear theory is applicable. To study the deviations from linear theory we consider an initially more nonlinear bending wave with initial warp amplitude equal to the aspect ratio of the disk at the inner boundary (and approximately 1/2 of the aspect ratio at warp center), and a thicker disk, for which aspect ratio at the center of the warp approximately equals 0.16. The details for these runs can be found in Table \ref{tabl:warp}.

Figure \ref{fig:warp_nonlin} shows the inclination angle in one snapshot at $t = 0.3 t_{\rm bend}$ for both of these runs.  It is clear from the left panel that the increase in wave amplitude does not show much effect on the shape of the curve. The agreement between the numerical results and the analytic estimates still very good. On the other hand, increasing the thickness of the disk (as in the right panel) leads to drastic change in shape of the curve, and to much worse agreement between the numerical and analytic solutions. In this case, the aspect ratio of the disk at the inner boundary is approximately 0.07, while the aspect ratio at the outer boundary is 0.21. The most probable cause of the difference between the numerical solution and analytic estimates are 3D effects due to the vertical structure of the disk. As we show next, same of these effects play an important role in disk-planet interactions as well.

\section{results: planet on a coplanar orbit in 3D}
\label{sect:coplanar}

\begin{figure*}
\includegraphics[width=\columnwidth]{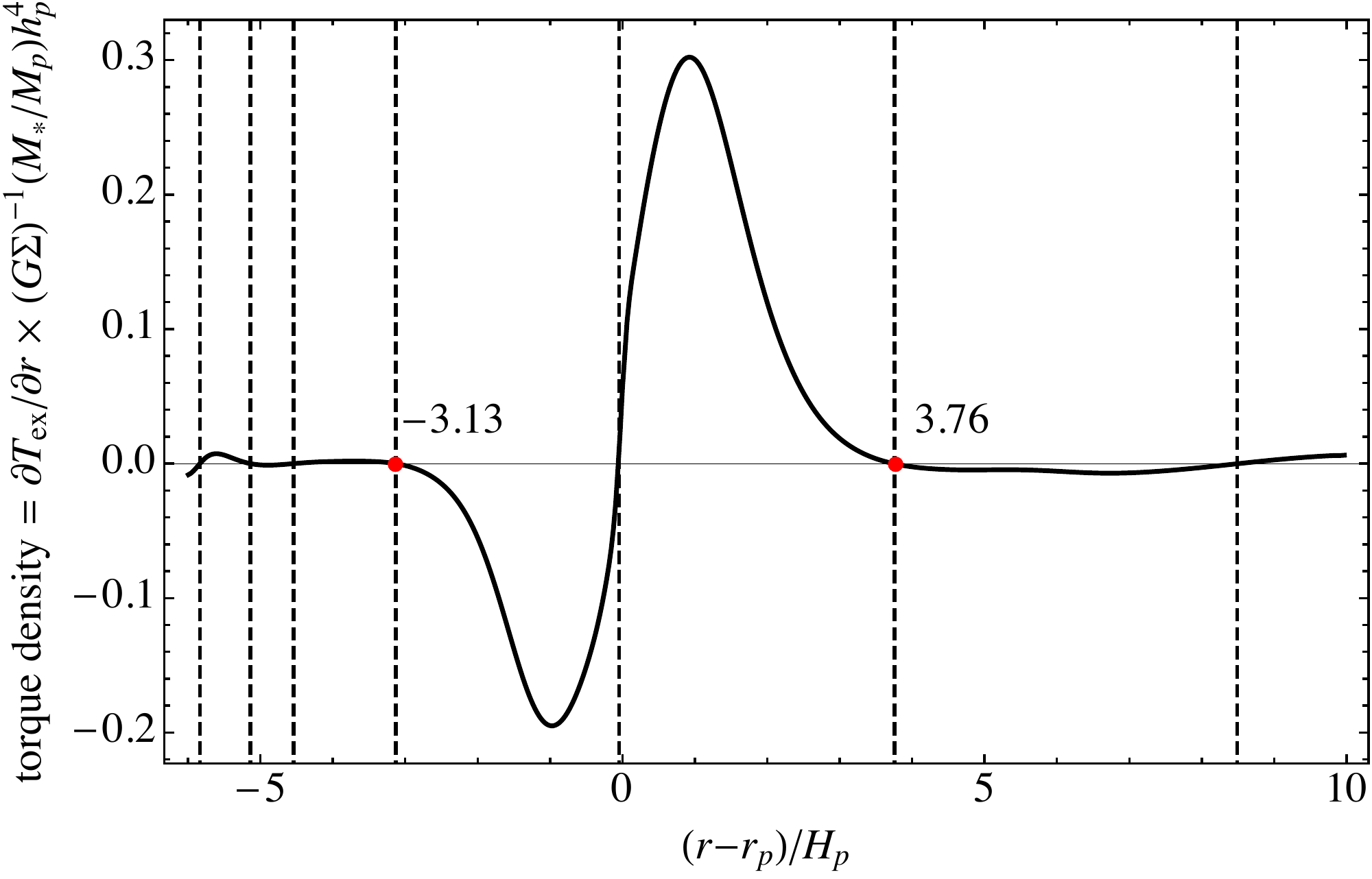}~~~~~\includegraphics[width=\columnwidth]{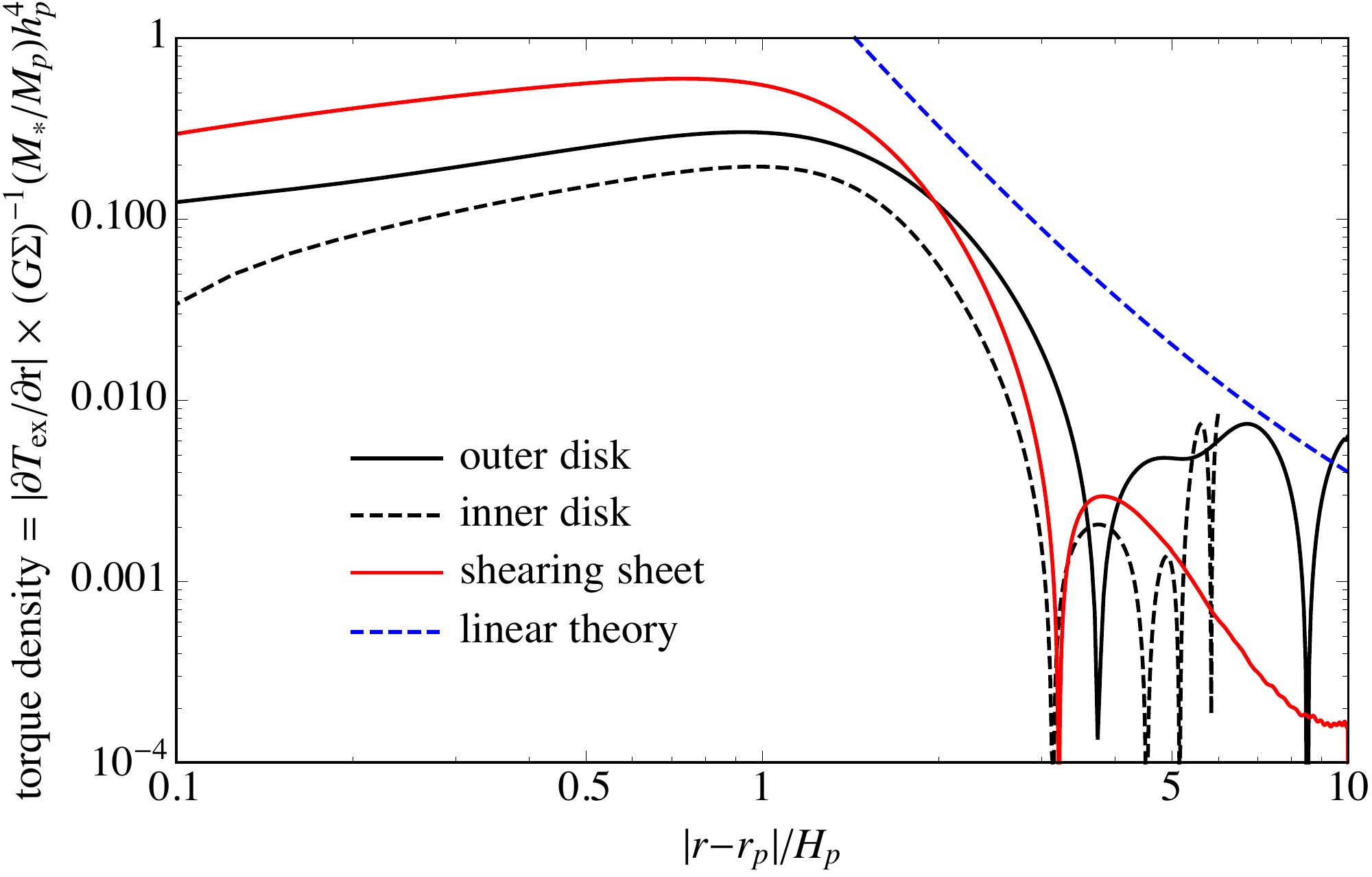}
\caption{\textbf{[Left panel]}: Torque density as a function of separation from planet $(r-r_{\rm p})/H_{\rm p}$ in full 3D.  Vertical dashed lines mark all the locations of the sign change of the torque.  The torque density is defined as an amount of angular momentum deposited to the disk, so it is mostly positive for $r>r_{\rm p}$ and mostly negative for $r<r_{\rm p}$. \textbf{[Right panel]}: Absolute value of the torque density from global 3D simulations (black lines) in comparison with 2D shearing sheet calculations \citep{2012ApJ...758...33P, 2012ApJ...747...24R} (red line), and linear theory \citet{GT80} (blue dashed line). The solid (dashed) black line corresponds to $r > r_{\rm p}$ ($r < r_{\rm p}$).}
\label{fig:Tex}
\end{figure*}

Before studying the interaction between a disk with a planet on an inclined orbit, it is instructive to start with the case of zero inclination. This case is widely studied in the literature. 
\citet{GT80} decomposed the potential of the planet into Fourier harmonics and obtained rather simple expression for the excitation torque density far from the planet:
\begin{align}
\label{eqn:Tex_GT80}
&\frac{\partial T_{\rm ex}}{\partial r} = {\rm sign}(r-r_{\rm p}) \Sigma \frac{G^2 M_{\rm p}^2}{(r-r_{\rm p})^4} \frac{\kappa^2 r}{8 A^4} C_{\rm GT80},\\
&C_{\rm GT80} =  \left\{ \frac{2 \Omega}{\kappa}K_0 \left(\frac{\kappa}{2 |A|} \right)+  K_1 \left(\frac{\kappa}{2 |A|} \right) \right\}^2.\nonumber
\end{align}
Here $G$ is the gravitational constant, $\Sigma$ is the surface density of the disk, and $\Omega(r)$, $\kappa(r)$ and $A(r) = \tfrac{1}{2}r \tfrac{{\rm d}\Omega}{{\rm d} r}$ are the angular velocity, epicyclic frequency and Toomre A parameter of the disk at distance $r$ from the primary star respectively.  Note that the we define the torque density as an amount of angular momentum excited in form of density waves, which implies that the total torque in the outer disk $r>r_{\rm p}$ is positive, while the total torque in the inner disk $r<r_{\rm p}$ is negative.

For a Keplerian disk, the expression (\ref{eqn:Tex_GT80}) takes an even simpler form
\begin{align}
\label{eq:2.5}
\frac{\partial T_{\rm ex}^{\rm GT80}}{\partial r} &= \frac{ \Sigma G^2 M_{\rm p}^2 r}{\Omega^2 H_{\rm p}^4} \psi_{\rm GT80}\left(\frac{r-r_{\rm p}}{H_{\rm p}}\right) ,\\
\label{eq:psi2.5}
 \psi_{\rm GT80} (x) &\equiv\frac{32}{81} \frac{{\rm sign}(x)}{x^4} \left\{2K_0 (2/3) +K_1(2/3)  \right\}^2 \approx \nonumber\\
 &\approx 2.5 \frac{{\rm sign}(x)}{x^4}, 
\end{align}
with $H_p$ equal to the unperturbed disk thickness near the perturber's orbit. This expression is valid at large separations from the planet $|r-r_{\rm p}|/H_{\rm p} \gg 1$. Near the planet the torque cuts off and approaches 0 at $r = r_{\rm p}$. 

The excitation of density waves near the planet was studied in detail by \citet{2012ApJ...747...24R} and \citet{2012ApJ...758...33P}  who noted the importance of overlapping Linblad resonances near the perturber.  Most notably, the solution in the 2D shearing sheet approximation has a negative torque at radial separation $|r-r_{\rm p}| \sim 3 h_{\rm p}$. 

The left panel of Figure \ref{fig:Tex} shows the torque density obtained from 3D isothermal hydrodynamic simulation of planet-disk interactions with $M_{\rm p } = 10^{-4} M_\star$, and $h_{\rm p} = H_{\rm p}/r_{\rm p} = 0.1$. The overall shape of the torque density is in very good agreement with previous numerical calculations \citep{2008ApJ...685..560D,2010ApJ...724..730D}. On the right panel of Figure \ref{fig:Tex} we compare the torque density with the results of numerical shearing sheet calculations in 2D \citep{2012ApJ...758...33P} as well as linear 2D predictions from \citet{GT80}. One can see that the form of the torque is in good agreement with shearing sheet estimates. The amplitude of the torque is $\sim 2-3$ times smaller in 3D reflecting the difference between density waves in 2D and 3D \citep[for more details, see][]{1993ApJ...419..166A}.

At a radial distance of several pressure scale heights from the planet the torque changes sign. \citet{2012ApJ...747...24R} found analytically that the sign alternation occurs at $r-r_{\rm p} \sim 3.2 H_{\rm p}$ in agreement with a number of numerical studies \citep{2003MNRAS.341..213B,2008ApJ...685..560D,2011ApJ...741...56D}.  The left panel of Figure \ref{fig:Tex} confirms the negative torque phenomenon. For the inner disk, $r < r_{\rm p}$, the torque changes sign at $|r - r_{\rm p}| \approx 3.13 H_{\rm p}$. On the other hand, in the outer disk $r > r_{\rm p}$, the sign alternation occurs at $|r - r_{\rm p}| \approx 3.76 H_{\rm p}$. Similar differences in the location of the change in sign between the inner and outer disk was also reported by \citet{2012ApJ...755....7D} in 2D. Figure \ref{fig:Tex} confirms the presence of the negative torque component for the first time in 3D. Note that in 2D torque changes sign only once. Our simulation domain allow us to see four different locations of sign change in the inner disk, and two such locations in the outer disk. This suggests that in a larger disk, there will be many sign changes of the torque. Although this phenomenon is interesting from the theoretical point of view, its contribution to the total torque is small and is unlikely to change the behavior of the disk significantly.

\section{results: planet on an inclined orbit}
\label{sect:inclined}

Before presenting the results of our numerical calculations, we first derive the evolution equations for the orbital elements of a planet moving on an inclined orbit. Following \citet{1976AmJPh..44..944B}, we define three components of the force
\begin{equation}
\boldsymbol{F} = R \boldsymbol{\hat e_R} + T \boldsymbol{\hat e_T} + N \boldsymbol{\hat e_N},
\end{equation}
where $\boldsymbol{\hat e_R}$, $\boldsymbol{\hat e_T}$ and $\boldsymbol{\hat e_N}$ are unit vectors along the radius vector to the planet position, velocity vector of the planet, and normal vector to the orbital plane of the planet respectively.

These components of the force correspond to the following evolution equations for the orbit \citep{1976AmJPh..44..944B,2011A&A...530A..41B}:
\begin{eqnarray}
    \frac{{\rm d} r_{\rm p}}{{\rm d} t} &=& 2~\frac{r_{\rm p}^{3/2}}{(GM_\star)^{1/2}} T,\\
    \frac{{\rm d} i_{\rm p}}{{\rm d} t} &=& -\frac{r_{\rm p}^{1/2}}{(GM_\star)^{1/2}} N \sin(\Omega_{\rm p} t),\\
    \frac{{\rm d} e_{\rm p}}{{\rm d} t} &=& ~~~\frac{r_{\rm p}^{1/2}}{(GM_\star)^{1/2}} [R \cos(\Omega_{\rm p} t) - 2 T \sin(\Omega_{\rm p} t)],\label{eq:ecc}
\end{eqnarray}
where we assume that the eccentricity is zero. Only components of the force lying in the orbital plane ($R$, $T$) can change the shape ($r_{\rm p}$, $e_{\rm p}$), while only component perpendicular to orbital plane ($N$) can change the inclination ($i_{\rm p}$) of the orbit.

In Sect.~\ref{sect:migration_inc} we focus on the $T$-component of the force and measure the migration timescale. In Sect.~\ref{sect:inclination_inc} and ~\ref{sect:precession_inc} we study the $N$-component of the force and study the inclination damping timescale as well as the rate of planetary orbital precession. Finally, in Sect.~\ref{sect:eccentricity_inc} we justify our assumption of a zero eccentricity orbit by calculating the rate of eccentricity excitation by the disk.

\subsection{Migration}
\label{sect:migration_inc}

\begin{figure*}
\includegraphics[width=\columnwidth]{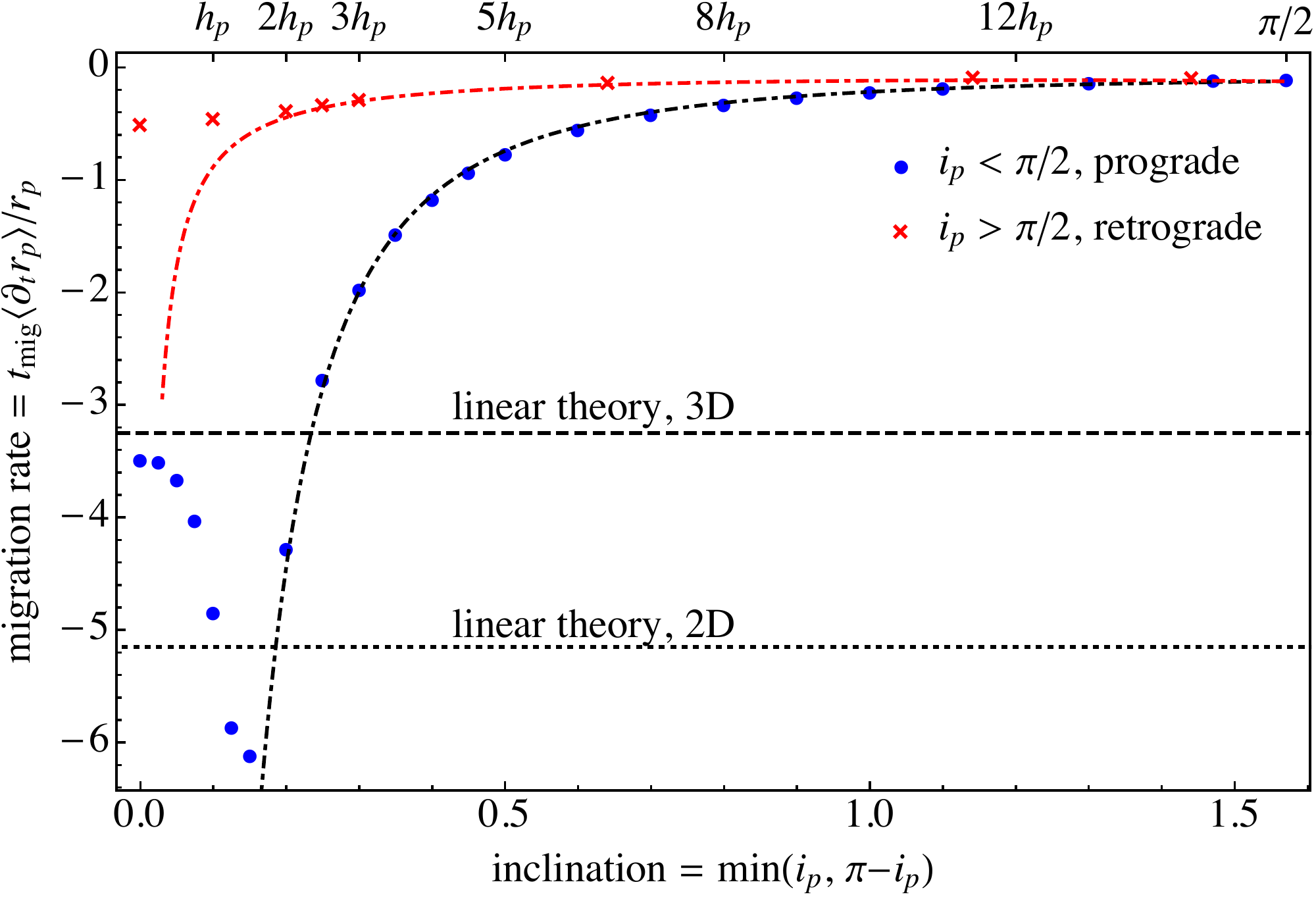}~~~~~\includegraphics[width=\columnwidth]{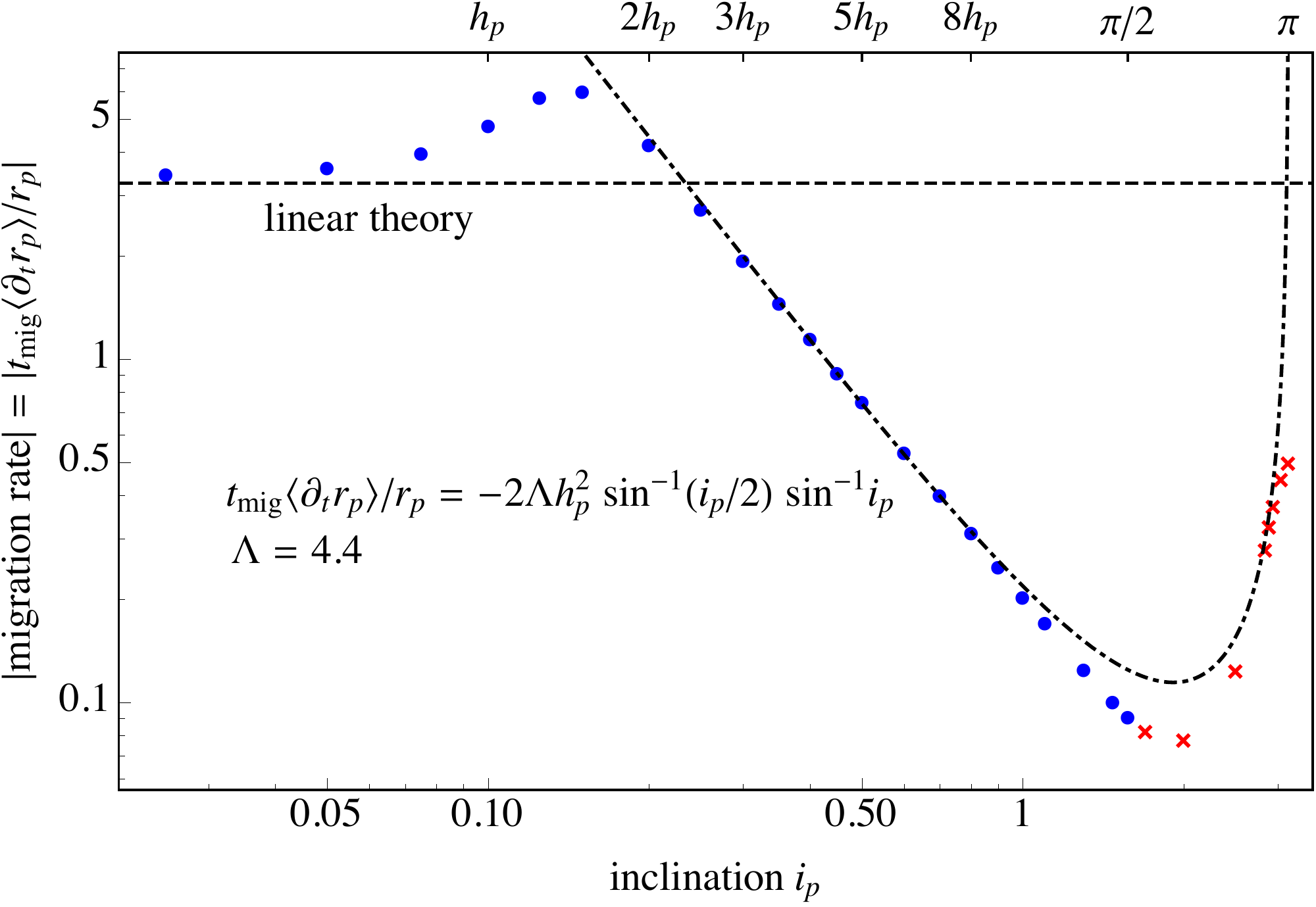}
\caption{\textbf{[Left panel]}: Planet migration rate as a function of inclination. Blue dots represent prograde orbits with inclination less than $\pi/2$, while red crosses represent retrograde orbits. Black dashed line corresponds to the results of linear theory \citep{2002ApJ...565.1257T} for a 3D disk with coplanar planet, equation (\ref{eqn:3dTTW}). The prediction for a 2D disk is shown with a black dotted line for comparison.
There is excellent agreement in the limit $i_{\rm p} \rightarrow 0$. At larger inclinations, the migration rate increases and reaches a maximum at $i_{\rm p} = 1.75 h_{\rm p}$.  At even larger inclinations, the migrations rate drops in good agreement with dynamical friction predictions of \citet{2012MNRAS.422.3611R}, equation \ref{eqn:rein12mig} (black dot-dashed line for $i_{\rm p} < \pi2$ and red dot-dashed line for $i_{\rm p}>\pi/2$).
\textbf{[Right panel]}: The same as left panel, but on a logarithmic scale.  The black dashed line shows the predictions of linear theory, while the black dot-dashed line shows the prediction from dynamical friction estimates.}
\label{fig:migration}
\end{figure*}

\begin{figure*}
\includegraphics[width=\columnwidth]{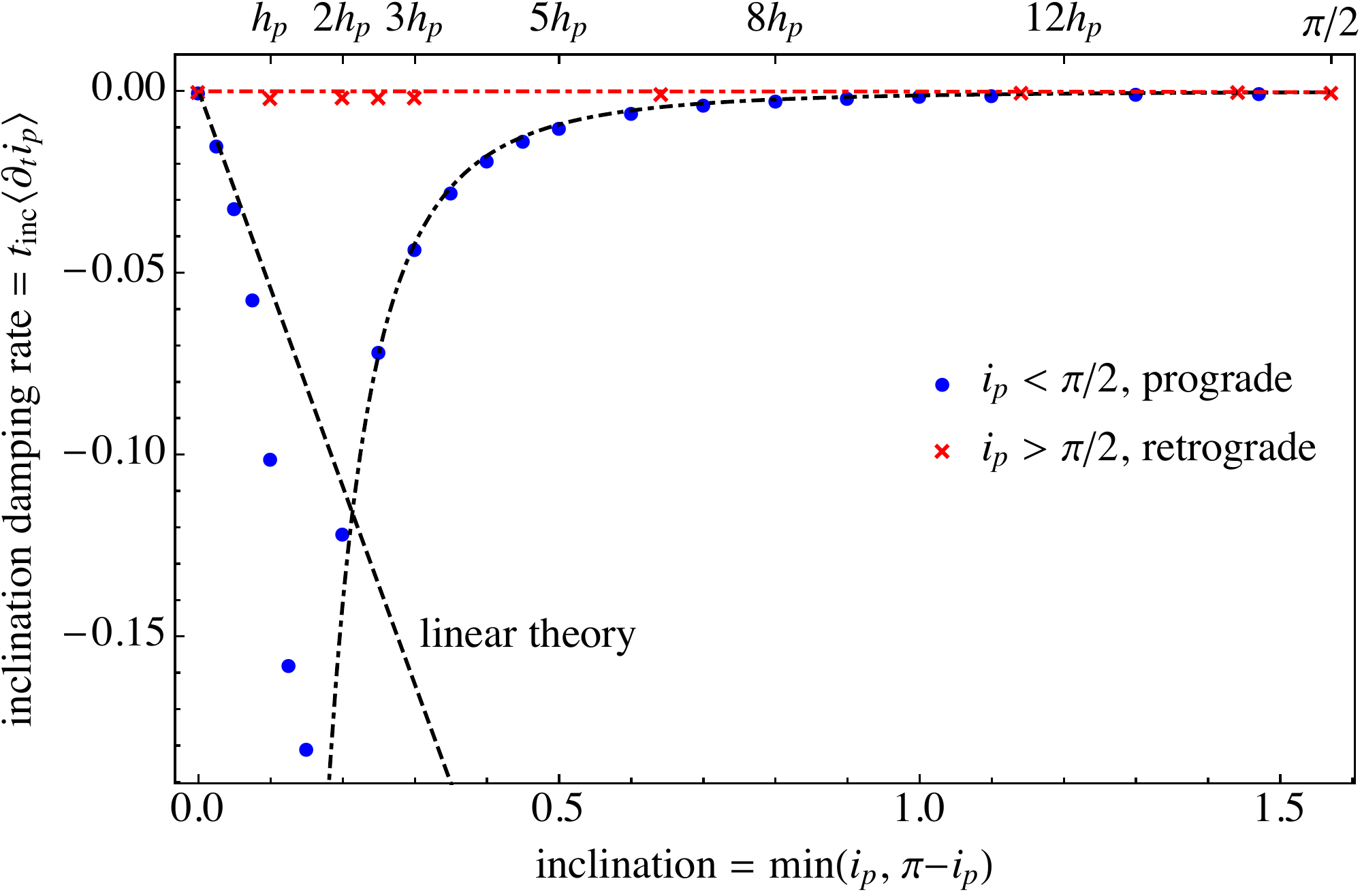}~~~~~\includegraphics[width=\columnwidth]{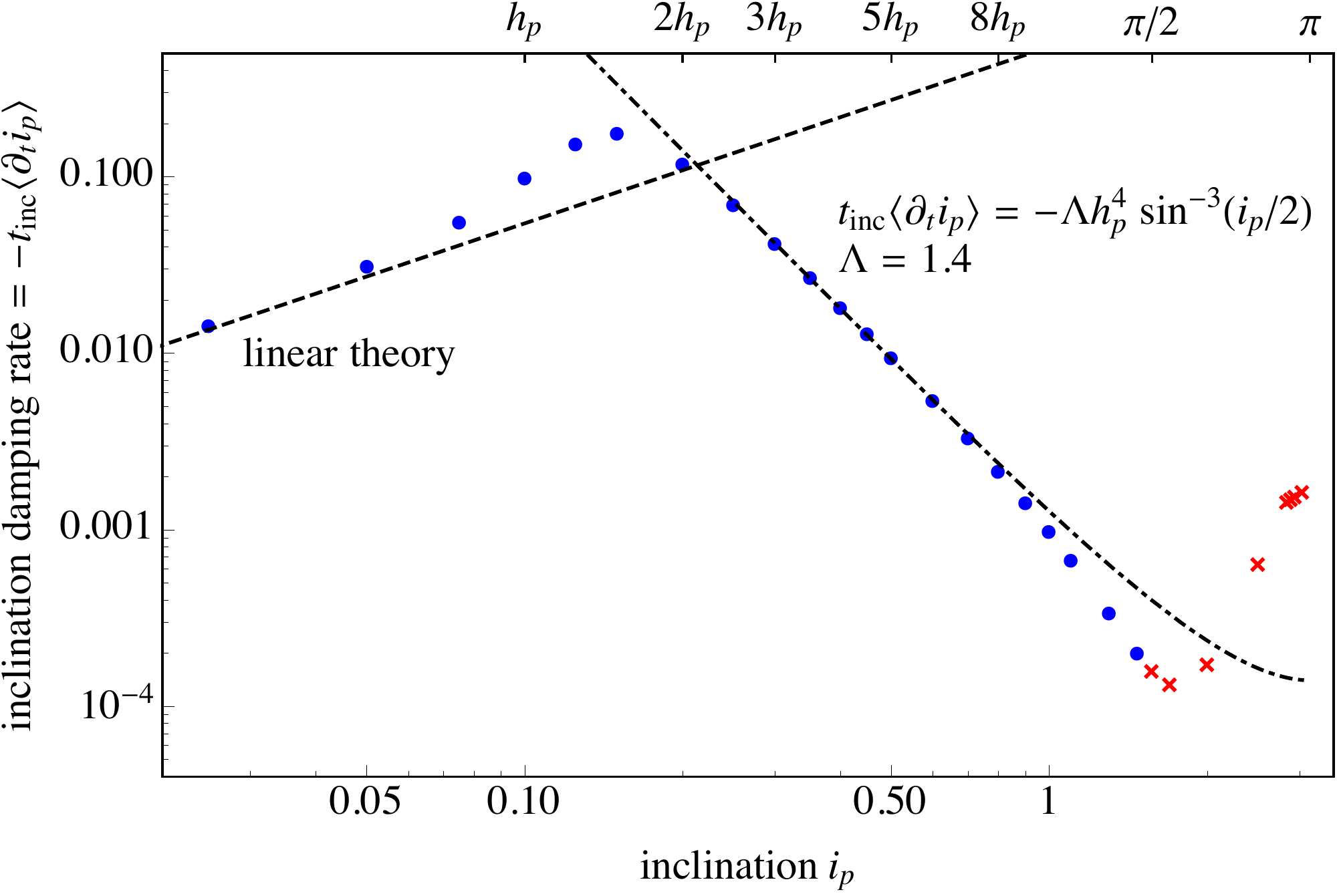}
\caption{\textbf{[Left panel]}: Inclination damping rate as a function of inclination. Dashed line corresponds to the linear shearing sheet calculations of \citet{2004ApJ...602..388T}, equation (\ref{eqn:TW04inc}), which fits our simulations very well at low inclinations. At intermediate inclinations, the damping rate increases and reaches its maximum at $i_{\rm p } = 1.75 h_{\rm p}$. At large inclination, the damping rate drops significantly. We find very good agreement with results of \citet{2012MNRAS.422.3611R}, equation (\ref{eqn:rein2012inc}), represented by the black (red) dot-dashed line for prograde (retrograde) orbits.
\textbf{[Right panel]}: The same as left panel, but on a logarithmic scale. Black dashed line shows the linear shearing sheet result, while black dot-dashed line shows the dynamical friction prediction.}
\label{fig:inclination}
\end{figure*}

In this section we derive an expression for the migration rate $\partial_t{ r}_{\rm p}/r_{\rm p}$ (where $\partial_t \equiv {\rm d}/{\rm d}t$), for use in comparison to our numerical results.
\citet{2002ApJ...565.1257T} found the following expression for the migration rate of a \emph{zero-inclination} planet:
\begin{align}
\label{eqn:3dTTW}
\frac{\left<\partial_t{r}_{\rm p}\right>}{r_{\rm p}} &= - (2.7 + 1.1 \alpha) \frac{M_{\rm p}}{M_\star} \frac{\Sigma_p r_{\rm p}^2}{M_\star} \left(\frac{H_{\rm p}}{r_{\rm p}} \right)^{-2} \Omega_{\rm p} = \\
&= - (2.7 + 1.1 \alpha) t_{\rm mig}^{-1}, \qquad {\rm (3D)}\nonumber
\end{align}
where the brackets $\left<...\right>$ are used for time averaging.  The migration timescale is
\begin{equation}
t_{\rm mig} = \Omega_{\rm p}^{-1} h_{\rm p}^{2}\left(\frac{M_{\rm p}}{M_\star}\right)^{-1} \left(\frac{\Sigma_p r_{\rm p}^2}{M_\star}\right)^{-1}.
\end{equation}
For our typical simulation with $M_{\rm p}/M_\star = 10^{-4}$ and $h_{\rm p} = 0.1$, the migration timescale is $t_{\rm mig} \approx 100~\Omega_{\rm p}^{-1} (\Sigma_{\rm p}r_{\rm p}^2 / M_\star)^{-1}$. The parameter $\alpha$ is the radial slope of surface density profile. Our isothermal disk has $\alpha = 0.5$.

Figure \ref{fig:migration} shows the net migration rate as a function of inclination.  The black dashed line show the analytic predictions of \citet{2002ApJ...565.1257T}. In addition to 3D expression (\ref{eqn:3dTTW}), we compare the migration rate with the expression for disk assuming 2D symmetry from \citet{2002ApJ...565.1257T} (shown as the black dotted line):
\begin{align}
\frac{\left<{\partial_t r}_{\rm p}\right>}{r_{\rm p}} &= - (2.32 + 5.66 \alpha) t_{\rm mig}^{-1}. \qquad {\rm (2D)}. \label{eqn:2dTTW}
\end{align}
Formally, the expressions of \citet{2002ApJ...565.1257T} are only applicable for zero inclination.  From Figure \ref{fig:migration}, in the limit of $i_{\rm p} \rightarrow 0$ the linear theory of \citet{2002ApJ...565.1257T} is in a very good agreement with the results of our simulations.

At moderately large inclinations, $i_{\rm p} \sim h_{\rm p}$ the torque acting on the planet is considerably larger, with the net migration rate increasing by a factor of 2, and reaching its maximum at $i_{\rm p} \sim 1.75 h_{\rm p}$.  The value at maximum is roughly two times larger than the predictions of linear theory.

For even larger inclinations, the migration rate drops dramatically. \citet{2012MNRAS.422.3611R} computed the rate of change of the orbital elements for very high inclination planets resulting from dynamical friction due to planet passage through the disk. They found the following expression for the net migration rate:
\begin{equation}
    \label{eqn:rein12mig}
    \left<\partial_t r_{\rm p}\right>/r_{\rm p} = - t_{\rm mig}^{-1}\cdot 2h_{\rm p}^2 \sin^{-1}(i_{\rm p}/2) \sin^{-1} i_{\rm p} \cdot \Lambda,
\end{equation}
where $\Lambda$ is a Coulomb logarithm, which depends on a number of parameters including the size of the disk and the smoothing length. As it is not well constrained, we use it as a fitting parameter for the high-inclination region of Figure~\ref{fig:migration}. We find that the dynamical friction prediction (\ref{eqn:rein12mig}) provides a very good fit to the results for intermediate inclination angles ($i_{\rm p} \la 1$), with $\Lambda = 4.39$ providing the best fit. However, for the largest inclinations, including retrograde orbits, this fit disagrees with simulations; we comment on this discrepancy at the end of the next subsection.

\subsection{Inclination damping}
\label{sect:inclination_inc}

\begin{figure*}
\includegraphics[width=\columnwidth]{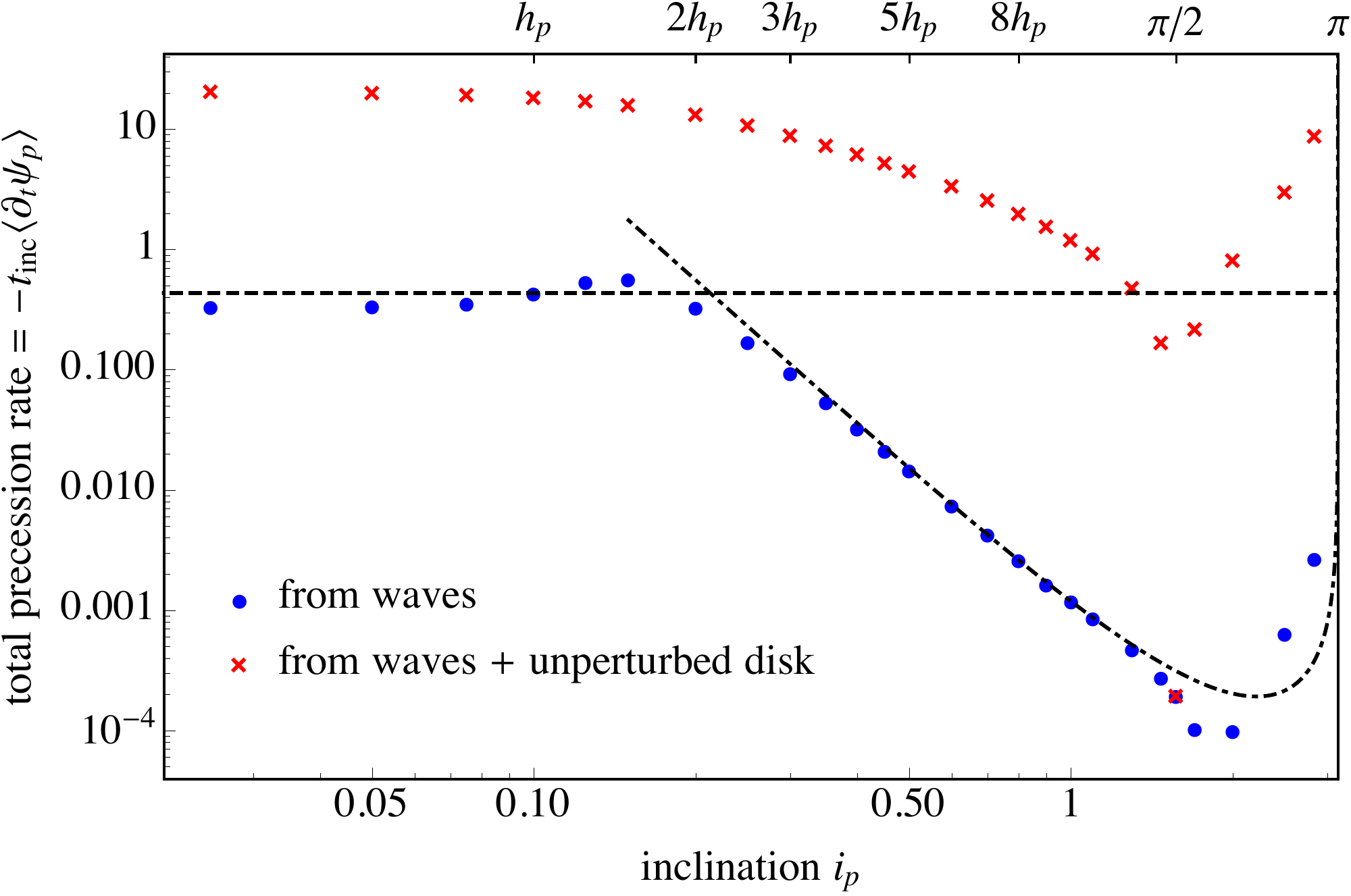}~~~~~\includegraphics[width=\columnwidth]{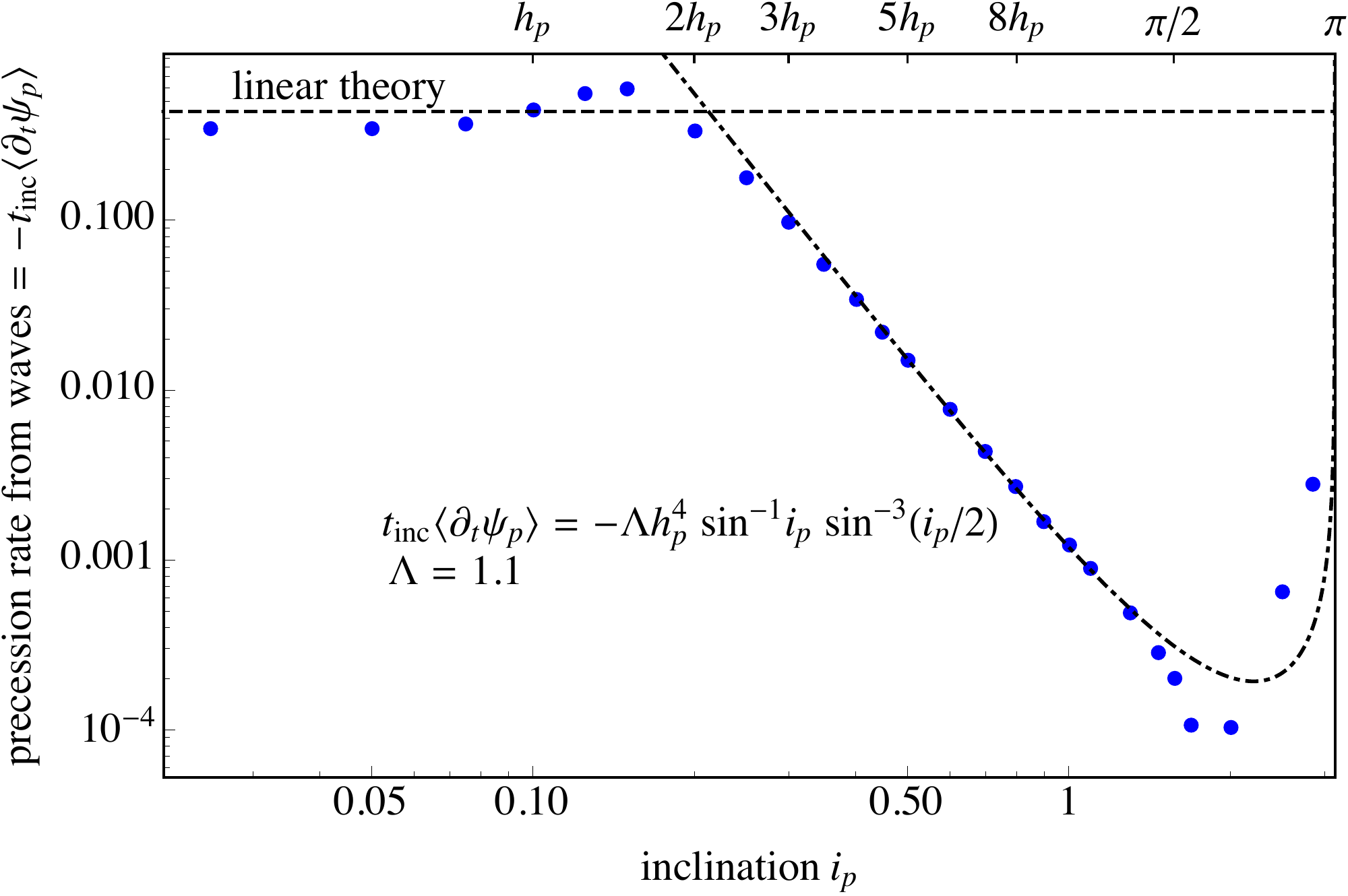}
\caption{\textbf{[Left panel]}: Precession of planetary orbit as a function of inclination. Red crosses corresponds to the secular effect from the unperturbed disk. Blue points correspond to the effect from density and bending waves. The secular effect is $\sim 100$ times larger than that from waves. 
\textbf{[Right panel]}: Precession rate driven by waves as compared to linear theory \citep{2004ApJ...602..388T}, equation (\ref{eqn:tw04prec}) dashed line, or dynamical friction \citep{2012MNRAS.422.3611R}, equation (\ref{eqn:rein2012prec}), dash-dot line. This comparison shows very good agreement and provides an additional test of the theory.}
\label{fig:precession}
\end{figure*}

\begin{figure}
\includegraphics[width=\columnwidth]{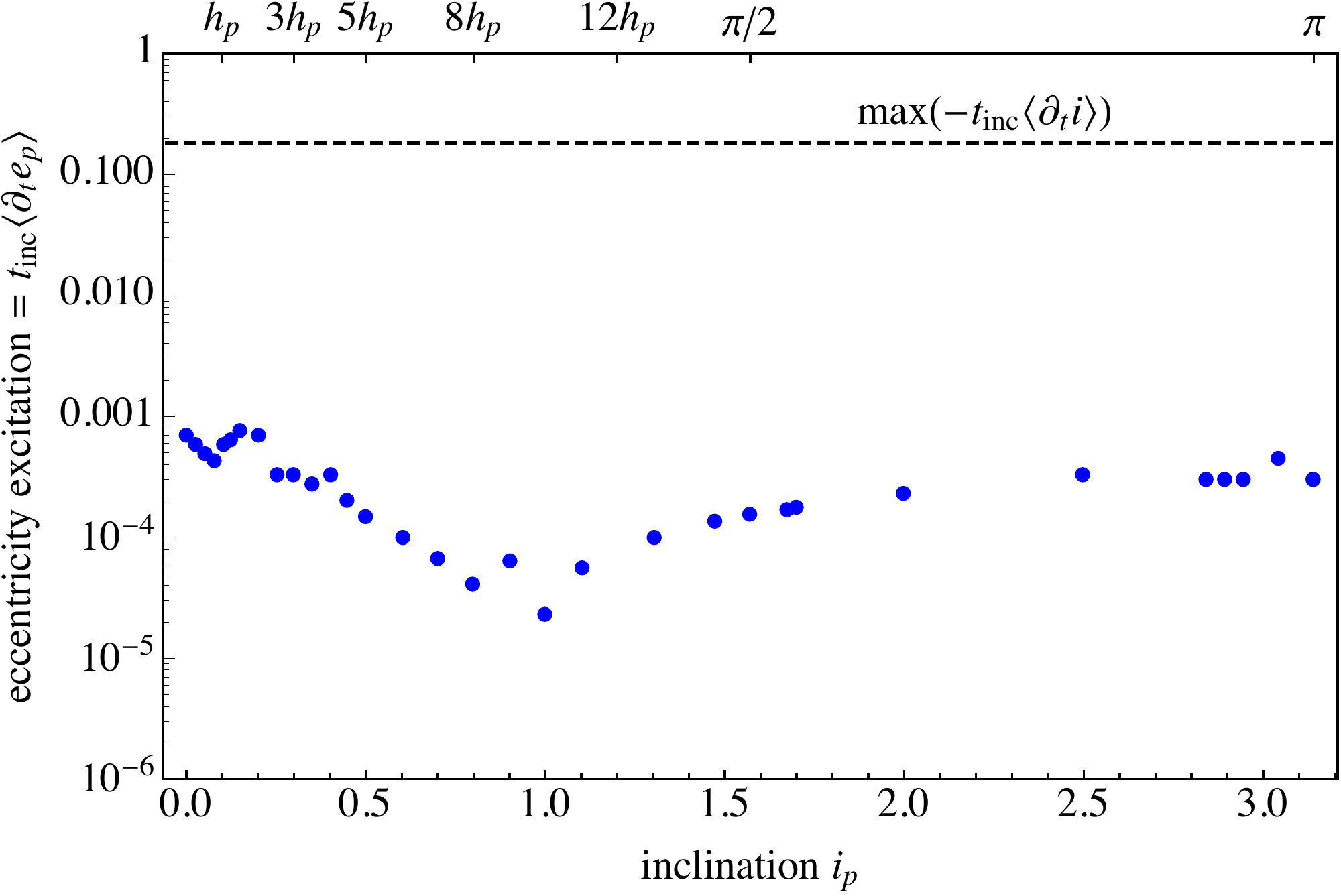}
\caption{Eccentricity excitation rate as a function of planetary inclination. The time is normalized by $t_{\rm inc}$, implying that the characteristic timescale of eccentricity excitation is $10^4 t_{\rm inc} = 10^{2} t_{\rm mig} \gg t_{\rm inc},t_{\rm mig}$. We thus conclude that initially circular orbits stay almost circular throughout orbital evolution.}
\label{fig:eccentricity}
\end{figure}

We now investigate the evolution of planetary inclination. For inclined planets, \citet{2004ApJ...602..388T} made analytical predictions for the damping of inclination:
\begin{align}
\label{eqn:TW04inc}
\frac{\left<{\partial_t i}_{\rm p}\right>}{i_{\rm p}} &= -0.544 \frac{M_{\rm p}}{M_\star} \frac{\Sigma_p r_{\rm p}^2}{M_\star} \left(\frac{H_{\rm p}}{r_{\rm p}} \right)^{-4} \Omega_{\rm p} = \\
& = -0.544 t_{\rm inc}^{-1} \nonumber,
\end{align}
where the inclination damping timescale $t_{\rm inc}$ is the same as $t_{\rm wave}$ from \citet{2004ApJ...602..388T}:
\begin{equation}
    t_{\rm inc} = \Omega_{\rm p}^{-1} h_{\rm p}^{4}\left(\frac{M_{\rm p}}{M_\star}\right)^{-1} \left(\frac{\Sigma_p r_{\rm p}^2}{M_\star}\right)^{-1}.
\end{equation}
For our typical simulation with $M_{\rm p}/M_{\star}=10^{-4}$ and $h_{\rm p}=0.1$, the inclination damping timescale is $t_{\rm inc} \approx \Omega_{\rm p}^{-1} (\Sigma_{\rm p}r_{\rm p}^2 / M_{\star})^{-1}$. This timescale $t_{\rm inc}$ is $h_{\rm p}^{-2}$ times shorter than $t_{\rm mig}$. This means the inclination damps much faster than the migration timescale (in our case, 100 times faster) for small inclinations.

One should note, however, that the linear calculations of \citet{2004ApJ...602..388T} assume that $i_{\rm p} \ll h_{\rm p}$ and are performed in the shearing sheet approximation. In our simulations, the smallest inclination we study is $i_{\rm p, min} = 0.25 h_{\rm p}$, so linear theory is not formally applicable to our results. 
Figure \ref{fig:inclination} shows the inclination damping rate as a function of inclination.  The black dashed line shows the comparison with linear theory \citet{2004ApJ...602..388T}. One can see that in the limit of small inclinations the agreement is very good. 

At intermediate inclinations $i_{\rm p}\sim h_{\rm p}$ inclination damping rate starts to deviate from linear predictions, and reaches it's maximum at $i_{\rm p}\sim 1.75 h_{\rm p}$. Interestingly, the maximum in inclination damping rate corresponds to the maximum in migration rate. One can conclude that at $i_{\rm p} \la 2 h_{\rm p}$ the damping is almost exponential with timescale similar to the predictions of \citet{2004ApJ...602..388T}.
At larger inclinations, the inclination damping rate drops significantly. \citet{2007A&A...473..329C} and \citet{2011A&A...530A..41B} predict that $\partial_t i_{\rm p} \propto i_{\rm p}^{-2}$. Our study, however, shows quite different behavior.   Fitting a power-law to our numerically measured damping rates gives an index which is closer to $-3$ rather than $-2$.

This disagreement can be easily explained.  Previous studies used a rather small range in values for the inclination, with the largest values used being in the transition region between linear and high-inclinations limits. For very high inclinations we can again use the results of \citet{2012MNRAS.422.3611R}. The inclination damping rate due to dynamical friction is
\begin{equation}
    \label{eqn:rein2012inc}
    \left<\partial_t i_{\rm p}\right> = - t_{\rm inc}^{-1}\cdot h_{\rm p}^4 \sin^{-3}(i_{\rm p}/2)\cdot \Lambda,
\end{equation}
where $\Lambda$ is again a Coulomb logarithm which we use as a free parameter for fitting the inclination damping curve. We conclude that this expression is in good agreement with simulations for $\Lambda = 1.46$ (again, only for intermediate inclinations $i_{\rm p} \la 1$).

We note, that this Coulomb logarithm is different from the one used to fit the migration rate curve. However, the results of \citet{2012MNRAS.422.3611R} are based on the approximations made in \citet{1999ApJ...513..252O} where it is assumed that waves propagate through a uniform medium. It is rather surprising that protoplanetary disks can be approximated as uniform medium to such a good accuracy.  It is likely that the disagreement between the analytic and numerical estimates at intermediate inclinations $i_{\rm p} \sim \pi/2$ results from the same reason why the best-fit Coulomb logarithm is different for different components of the torque: the protoplanetary disk is anisotropic, and this should be taken into account in dynamical friction calculations. One of the consequences of anisotropy might be the misalignment between the dynamical friction force and relative velocity between the planet and the disk fluid -- the assumption usually used in analytical calculations of dynamical friction.

Perfectly retrograde orbit with $i_{\rm p} = \pi$ is an unstable equilibrium: the inclination damping rate is exactly zero at $i_{\rm p} = \pi$, but it has a finite limit as $i_{\rm p}$ approaches $\pi$ meaning that the orbit becomes more polar. This property of inclination damping rate is seen from analytical prediction (\ref{eqn:rein2012inc}) as well as in numerical simulations \citep{2013MNRAS.431.1320X} in the context of massive planets. For such large inclination, the dominant force comes from dynamical friction which acts to reduce the misalignment.

\subsection{Precession}
\label{sect:precession_inc}

\begin{figure*}
\includegraphics[width=\columnwidth]{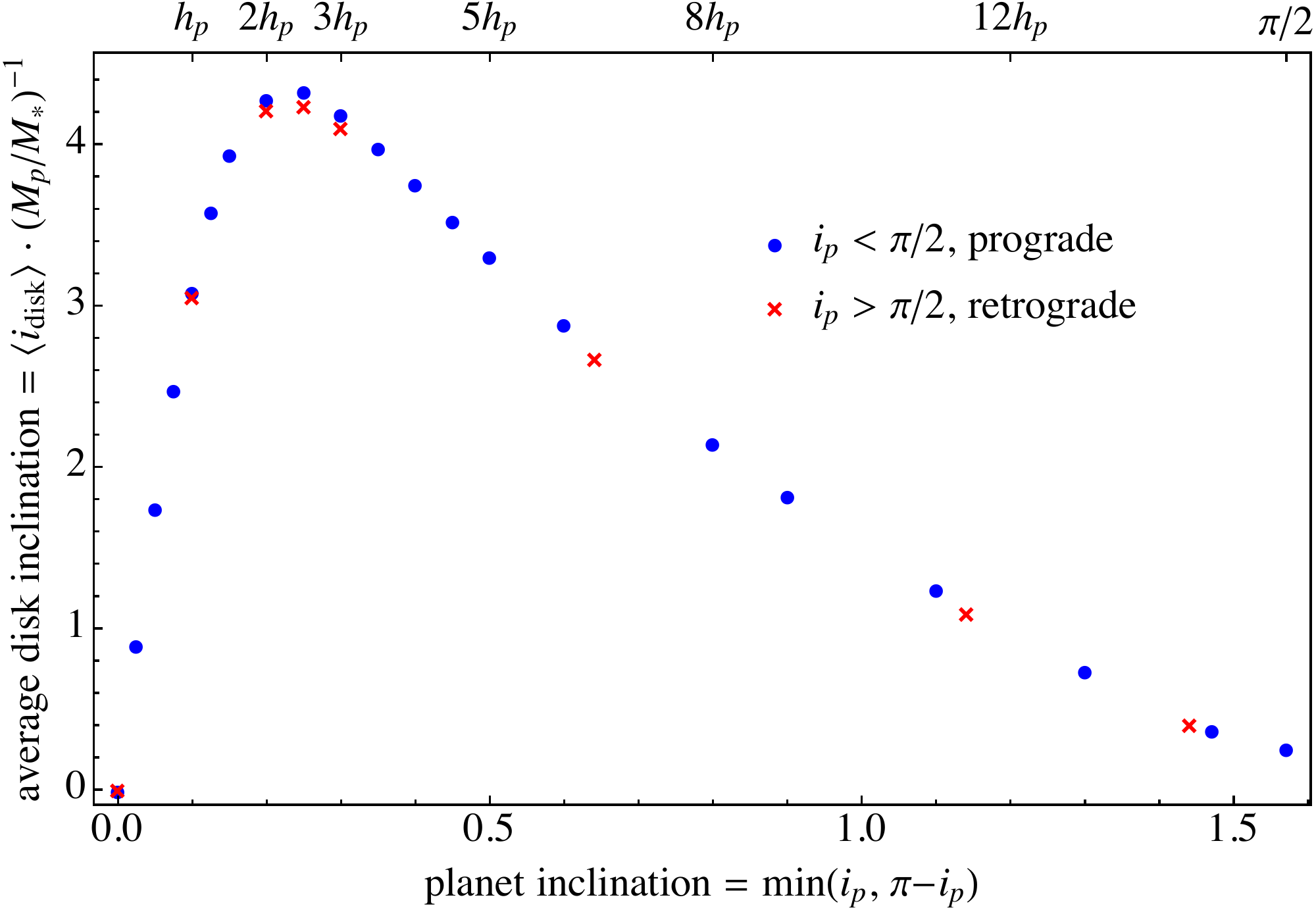}~~~\includegraphics[width=1.02\columnwidth]{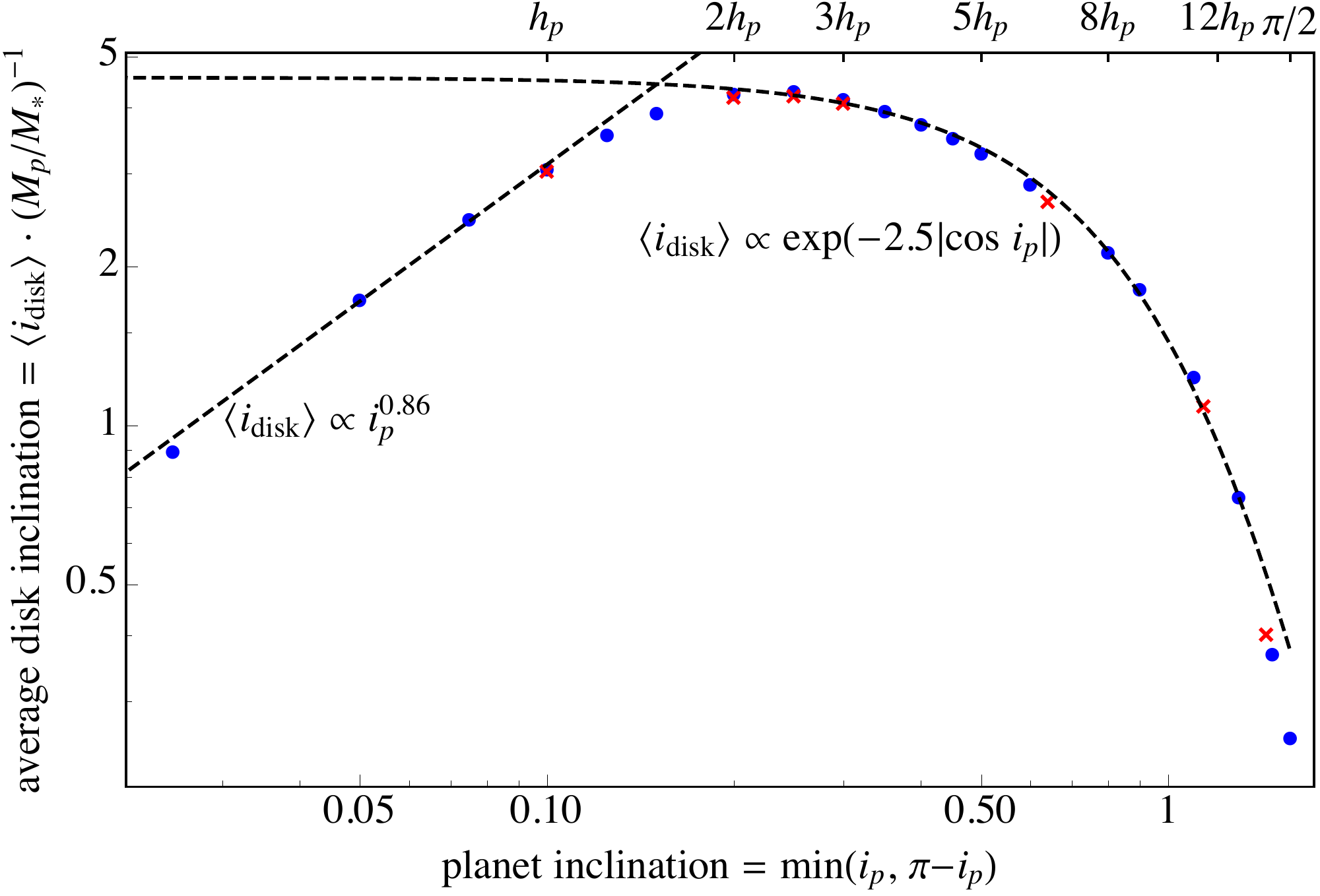}
\caption{\textbf{[Left panel]}: The angular momentum weighted inclination angle of the disk as a function of inclination of the planet. \textbf{[Right panel]}: The same as left panel, but in logarithmic scale, showing approximate scalings at small and large inclinations.}
\label{fig:response}
\end{figure*}

Next we study the planetary orbit precession rate $\partial_t \psi_{\rm p}$, given by
\begin{equation}
    \frac{{\rm d} \psi_{\rm p}}{{\rm d} t} \equiv \frac{r_{\rm p}^{1/2}}{(GM_\star)^{1/2}} N \frac{\cos(\Omega_{\rm p} t)}{\sin i_{\rm p}}.
\end{equation}
\citet{2004ApJ...602..388T} provide the linear prediction for this quantity
\begin{align}
    \label{eqn:tw04prec}
   \left< \partial_t \psi_{\rm p}\right> &= -0.435 \frac{M_{\rm p}}{M_\star} \frac{\Sigma_p r_{\rm p}^2}{M_\star} \left(\frac{H_{\rm p}}{r_{\rm p}} \right)^{-4} \Omega_{\rm p} = \\
& = -0.435 t_{\rm inc}^{-1} \nonumber.
\end{align}
This expression corresponds to the precession rate \emph{due to waves}. However, the unperturbed disk itself causes the orbit to precess. This secular effect is $h_{\rm p}^{-2} \approx 100$ times larger than the effect from perturbations.

In Figure \ref{fig:precession} we show the precession rate of the planetary orbit. Red points in the left panel correspond to the total precession rate, which consists of contributions from the unperturbed disk and waves. The former depends only on initial conditions and can be computed analytically. The blue dots in the left panel show the precession rate due to waves only (computed by subtracting the contribution from the unperturbed disk from the total rate).  One can see that the contribution from the unperturbed disk is dominant in determining the precession rate. 

From our simulations, it is possible to infer the impact of waves on precession rate, and compare it with analytical predictions. For small inclinations $i_{\rm p} \la h_{\rm p}$, the precession rate is almost constant with value close to linear estimate (\ref{eqn:tw04prec}). 
For larger inclinations, linear theory does not work very well.  We fit the large-inclination part of the precession rate with 
\begin{equation}
\label{eqn:rein2012prec}
    \left<\partial_t \psi_{\rm p}\right> = - t_{\rm inc}^{-1} \cdot h_{\rm p}^4\sin^{-1} i_{\rm p} \sin^{-3} (i_{\rm p}/2) \cdot \Lambda.
\end{equation}
The best fit parameter $\Lambda$ is 1.02. Note that equation (\ref{eqn:rein2012prec}) is not present in \citet{2012MNRAS.422.3611R}. The functional form of equation (\ref{eqn:rein2012prec}) just comes from the assumption that the inclination angle evolves on the timescale similar to the timescale of precession.

As mentioned above, waves are not important for precession rate of planetary orbit. However, they provide one more quantity which can be used to test the theory.

\subsection{Eccentricity excitation}
\label{sect:eccentricity_inc}

Throughout this paper we have assumed circular orbits. This assumption would break down if there is a large eccentricity excitation rate. 
In Figure \ref{fig:eccentricity} we show the eccentricity excitation rate $\partial_t e_{\rm p}$ measured from our simulations (equation \ref{eq:ecc}). One can see that the characteristic timescale of eccentricity excitation is 
\begin{equation}
t_{\rm ecc} \sim 10^{ 4} t_{\rm inc} = 10^{2} t_{\rm mig}.
\end{equation}
Thus we conclude that the eccentricity excitation rate is small, and our assumption of circular orbits is justified.

\section{Response of the disk and Observational signatures}
\label{sect:obs}

The interaction between a disk and an inclined planet induces spiral structure in the disk.
In addition, the disk may become warped and develop non-zero inclination. Figure \ref{fig:response} shows the inclination angle of the disk as a function of the inclination of the planet. To compute the former, we first compute the inclination of individual ``rings'' in the disk (using equation \ref{eq:delta(r)}), and then average over the disk using the value of the angular momentum in each rings as a weight. It is easy to see that the response of the disk peaks at $i_{\rm p} \sim 3 h_{\rm p}$. 

\begin{figure*}
\includegraphics[width=2.2\columnwidth]{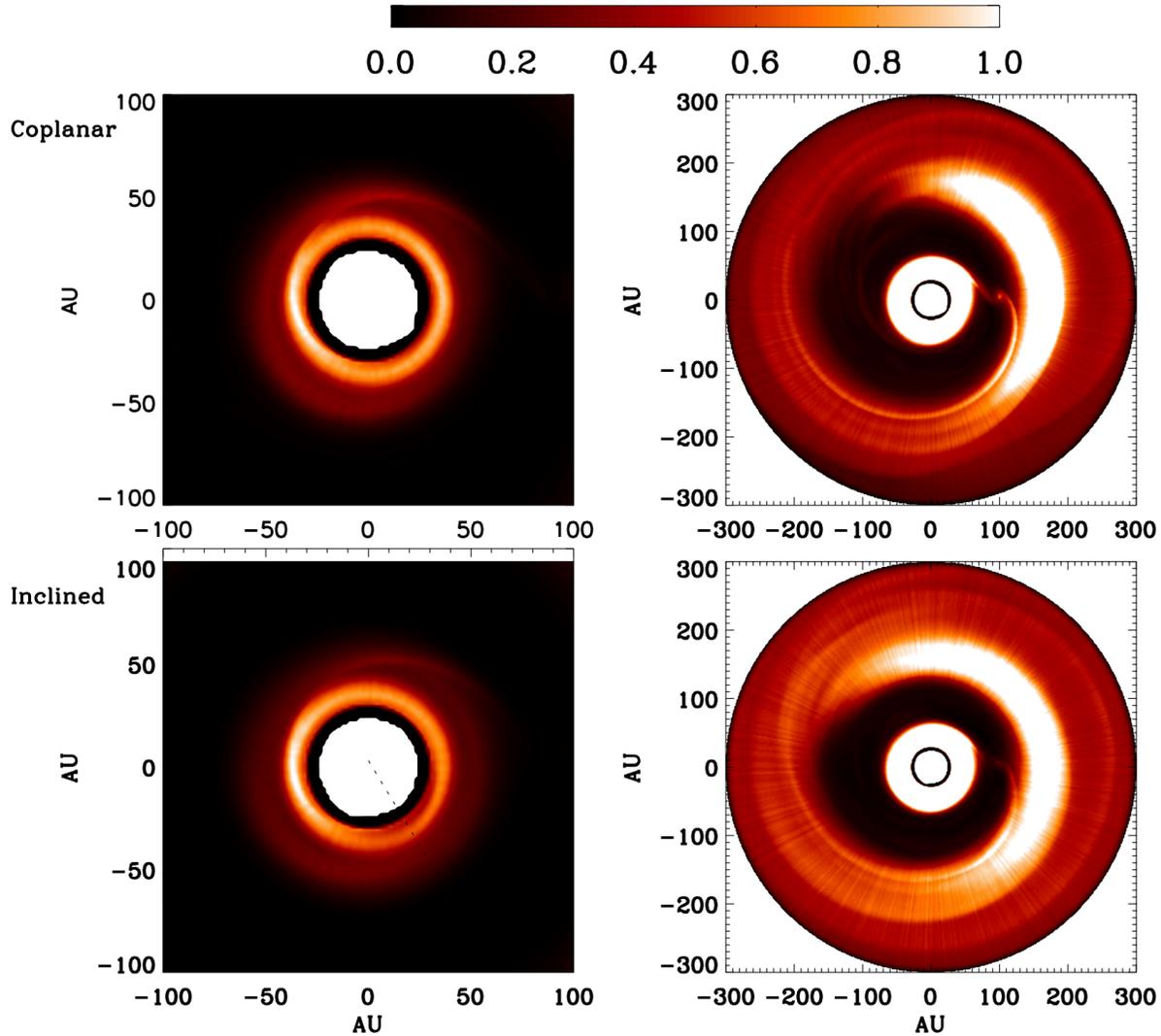}
\caption{ Polarized intensity images of disk with a coplanar (upper panels) and an inclined planet (lower panels). The images to the left show the inner parts of the disk, while images to the right show the full disk. Note that intensities on the images to the right are scaled to $1/10$ of the maximum value to emphasize the outer parts of the disk. The inner disk of the inclined planet shows clear asymmetries caused by the disk warp (indicated by dashed line). The inner disk for coplanar case does not have such asymmetry. The outer disk features asymmetries in both cases due to gap edge vortex. There are also additional spiral arms caused by the vortex.
}
\label{fig:twodg}
\end{figure*}

To understand the observed spiral pattern and to investigate whether it can be used to constrain the inclination of the planet with respect to the disk, we have carried out two additional simulations: one with a coplanar and one with an inclined planet.  From both simulations we have generated near-IR scattered light images, as described below. These two simulations have $H_{\rm p}/r_{\rm p}=0.1$ and $T(r)\propto r^{-1/2}$, values which differ from those used for the global isothermal simulations described above.  To mimic radiative cooling
processes in the disk we implement simple cooling prescription of
\citet{2015ApJ...813...88Z}. As stated in Equation 9 of that paper, the cooling timescale is $10^5$ orbital periods at 1 AU and $10^{-2}$
orbital periods at 100 AU. So we choose an intermediate cooling timescale value $T_{\rm cool} = 1/\Omega_{\rm p}$, which is also the value adopted
in that paper. 
The simulation domain is from $0.3 r_{\rm p}$ to $3 r_{\rm p}$ in the $r$-direction, [0.97,2.17] in the $\theta$-direction, and 2$\pi$ in the $\phi$-direction with the resolution of 256$\times$125$\times$686. The planet has mass $M_{\rm p} = 10^{-3} M_{\star}$ and a semi-major axis of $r_{\rm p}$ in both simulations. In the inclined planet case, the planetary orbital plane is inclined at 17 degrees ($\approx 3 H_{\rm p}/r_{\rm p}$). As shown above, this value is most effective at bending the disk. Note that the mass of the planet used in this section is significantly larger than in previous sections. We use such a heavy planet to enhance the appearance of the spiral arms. However, $M_{\rm p} = 10^{-3} M_\star $ is equal to the thermal mass $M_{\rm th} \equiv c_{\rm s}^3/G\Omega_{\rm p} = (H_{\rm p}/r_{\rm p})^3 M_\star$. This make the planet-disk interaction nonlinear, thus strictly speaking the analyses of previous sections does not apply.  Such massive planet also starts opening the gap. As we do not have any explicit viscosity, the gap continues to get deeper with time, and it is not fully formed during the time of our simulation. The spiral arms, however, are formed on the sound crossing time, which is much shorter than our simulation time.

We use RADMC-3D \footnote{\url{http://www.ita.uni-heidelberg.de/~dullemond/software/radmc-3d/}} for the 3D radiation transfer calculation. We use a spherical mesh that is the same as in our hydrodynamical simulations. To scale the quantities in our simulations, we assume the length unit is $r_{\rm p}=100$ AU, and the disk gas surface density at 100 AU is 1 g cm$^{-2}$ in the initial condition. The central star is a Herbig Ae star with $T_{\rm eff}=6810$ K, $R_{\star}=1.4 R_{\odot}$, and $M_{\star}=1.7 M_{\odot}$. We also assume a dust to gas mass ratio of 1/100 with 10\% of the dust mass in small grains which scatters the near-IR light (the same procedure as used in \citealt{2015ApJ...809L...5D}). The dust opacity is calculated using Mie theory for Magnesium-iron silicates \citep{1995A&A...300..503D}. The size of the small grains is 0.1 micron (the same as that in \citealt{2015MNRAS.451.1147J}).

The radiation transfer calculation has two steps. First, the temperature of the dust is determined. Then scattered light images are calculated including the full treatment of polarization. Full Stokes vectors $(I,Q,U,V)$ have been computed and the polarized intensity ($\sqrt{Q^2+U^2}$) images are shown in Figure \ref{fig:twodg}.  Images to the left show the intensity maps of the inner disk for coplanar (upper image) and inclined (lower image) planets. Polarized intensity is scaled to its maximum value. The bright region at the inner edge of the simulation domain on both images is due to the spiral arms. The intensity map of very inner disk in inclined case features some asymmetry (which we indicate by dashed line). Such asymmetry is not seen in coplanar case, and it is due to the slight warp of the disk. Images to the right show the full disk. To emphasize the outer parts of the disk, the intensity is scaled to $1/10$ of its maximum value. One can see clear asymmetries in the outer disk for both coplanar and inclined cases cause by the gap edge vortex. Although the gap edge vortex has been seen in many previous simulations with a coplanar planet \citep[e.g.,][]{2007A&A...471.1043D}, we have shown that
the gap edge vortex also exists even if the planet is on an inclined orbit. Furthermore, outer disk also has other spiral
features which are excited by vortex and are connected to it. We suggested that these spiral arms excited by the vortex can also be seen in the near-IR scattered light images.

\section{Conclusions}
\label{sect:conclusion}

We have conducted a series of high-resolution hydrodynamical numerical simulations of disk-satellite interaction. Our standard setup consists of a low-mass perturber with $M_{\rm p}/M_\star = 10^{-4}$ and three-dimensional isothermal disk with $h_{\rm p} = 0.1$. Perturbers have inclinations spanning from $i_{\rm p} = 0.25 h_{\rm p}$ to very large inclinations and retrograde orbits.

The study of coplanar planets allowed us to obtain the excitation torque density of the density waves (see Figure \ref{fig:Tex}). The migration rate from this torque is consistent with the analytic estimate of \citet{2002ApJ...565.1257T}. We confirm the interesting phenomenon of a negative torque density far from a planet \citep{2012ApJ...747...24R, 2012ApJ...758...33P} for the first time in 3D. For the inner disk, the first change in torque sign occurs at $|r-r_{\rm p}| \approx 3.13 H_{\rm p}$, while for the outer disk the change in sign happens at $|r-r_{\rm p}| \approx 3.76 H_{\rm p}$. Notably, even farther from the perturber, the torque changes sign again. Our simulation domain allows us to see 4 sign alternations suggesting that in a larger disk there will be many sign changes of the torque. These changes, however, do not contribute much to the total torque, and thus do not influence migration rates.  Overall, the torque density remains close to \citet{GT80} prediction, although the actual functional form is different at $|r-r_{\rm p}|\gg H_{\rm p}$ (see Figure \ref{fig:Tex}).

Simulations of inclined planets revealed the evolution of planetary orbital elements, namely semimajor axis, inclination angle, precession angle, and eccentricity (we consider initially circular orbits, and Figure \ref{fig:eccentricity} confirms that significant eccentricity excitation does not occur in the system).

The evolution of the semimajor axis (see Figure \ref{fig:migration}) leads to migration of planets with small inclination, and follows the results of linear calculations of \citet{2002ApJ...565.1257T} in the limit of small inclinations. For larger inclinations, the migration rate increases significantly, and peaks at inclination of $i_{\rm p} \sim 2h_{\rm p}$. The maximum migration rate is roughly 2 times larger that the value at zero inclination. For even larger inclinations, the migration rate drops and approximately follows the expression from \citet{2012MNRAS.422.3611R} for dynamical friction driven migration. There is a deviation from this expression occurring at inclinations close to $\pi/2$, as well as for retrograde orbits, but it holds for all intermediate inclinations $i_{\rm p} \la 1$.

In the limit of small inclinations, the inclination damping rate (see Figure \ref{fig:inclination}) approximately follows the linear predictions of \citet{2004ApJ...602..388T}. The inclination decays exponentially on a short timescale.  For inclinations up to $i_{\rm p} \sim 2h_{\rm p}$ the damping rate increases, although faster than predicted by \citet{2004ApJ...602..388T}. The maximum damping rate is roughly 2 times larger than predicted by linear theory. For larger inclinations, the damping rate drops.  Again, it approximately follows predictions based on dynamical friction \citet{2012MNRAS.422.3611R}, and deviates from these predictions only for very large inclinations and retrograde orbits. Our findings are in contradictions with the results of \citet{2007A&A...473..329C,2011A&A...530A..41B} who predicted $\partial_t i_{\rm p} \propto i_{\rm p}^{-2}$. For the intermediate inclinations, we found $\partial_t i_{\rm p} \appropto i_{\rm p}^{-3}$. In our simulations we used a much larger range of initial inclinations than in these previous works. In fact, \citet{2007A&A...473..329C} and \citet{2011A&A...530A..41B} only consider $i_{\rm p} < 15^\circ$, and have very few points in the limit $i_{\rm p} > 2 h_{\rm p}$.  We note that the numerical setup of \citet{2007A&A...473..329C,2011A&A...530A..41B} is different from the one we use in this paper. They use different disk scale height as well as finite viscosity. These differences may be the reason for contradiction between our studies, and future work with larger parameter study is needed to resolve this disagreement.

The study of the precession rate (see Figure \ref{fig:precession}) confirms the linear theory calculations \citep{2004ApJ...602..388T} in the limit of small inclinations, and dynamical friction predictions for intermediate inclinations. However, this statement only applies to the contribution from waves to the precession rate.   In fact, there is another much larger contributor, namely the unperturbed disk which is more important than waves on the precession rate for all inclinations except $\pi/2$ where the secular effects goes to zero.

As a result, Figures \ref{fig:migration}, \ref{fig:inclination}, and \ref{fig:precession} provide the evolution of the full set orbital elements as a function of inclination. For small inclinations, the linear theory holds. The torques peak at $i_{\rm p} \sim 2 h_{\rm p}$ with values roughly 2 times larger than linear predictions. For intermediate inclinations, dynamical friction calculations are approximately valid. Notably, the value of the best-fit Coulomb logarithm is different for different components of the torque, as one would expect due to the lack of isotropy in the system. This indicates that in such inhomogeneous systems the dynamical friction is not parallel to the perturber's velocity.

The interaction between an inclined planet and the disk causes the disk to warp and develop inclination (see Figure \ref{fig:response}). Because of that the scattered light images (Figure \ref{fig:twodg}) show asymmetries in polarized intensity caused by disk warp.
The time-dependent planetary potential produces this time-dependent spiral pattern. The period of variations of the pattern is equals to one half of the planetary orbit.

The inviscid hydrodynamical prescription adopted in this study may be oversimplified for the case of a more realistic protoplanetary disk. However, we do not expect a more realistic, e.g. non-ideal MHD, model of the gas dynamics of the disk to change our results dramatically, since if the planet only interacts with the disk through the gravitational torques the magnetic field will not change the
excitation torque density.  MHD effects might change the transfer of angular momentum from the waves to the disk, however this is not the focus of our paper.  

Our work is limited only to planets with small masses embedded in a thin disk. We expect our results for $M_{\rm p}/M_\star = 10^{-4}$ to apply for lighter planets. Heavier planets, on the other hand, quickly open a gap, which modifies the excitation of waves significantly. The disk-planet interaction in such systems is different for the inner and outer regions of the disk \citep{2017AJ....153...60M,2017arXiv170309250O}. Although the presence of inclination damping has been found for massive planets \citep{2013A&A...555A.124B,2017A&A...598A..70S}, we do not expect our torque expressions to hold in this regime.

The full description of planet-disk interaction requires understanding not only the excitation of density and bending waves, but also the mechanism of wave dissipation, and the resulting transfer of angular momentum from the waves to the disk \citep{2001ApJ...552..793G,2002ApJ...569..997R,2016ApJ...831..122R,Ju,Ju2017,2017arXiv171001304A}. This paper has studied only the excitation and propagation of such waves, not their damping.  We leave this subject for later studies.

\section*{ACKNOWLEDGEMENTS}

We thank Roman Rafikov, Ruth Murray-Clay, and Scott Tremaine for helpful discussions and comments. Z.Z. acknowledges support from the National Aeronautics and Space Administration through the Astrophysics Theory Program with Grant No. NNX17AK40G. 
All hydrodynamical simulations
were carried out using computers supported by the Princeton
Institute of Computational Science and Engineering and the Texas Advanced Computing Center (TACC) at The University of Texas at
Austin through XSEDE grant TG- AST130002.

\bibliographystyle{mn2e}
\bibliography{mybib}

\begin{thebibliography}{}

\bibitem[\protect\citeauthoryear{{Artymowicz}}{{Artymowicz}}{1993}]{1993ApJ...419..166A}
{Artymowicz} P.,  1993, \apj, 419, 166

\bibitem[\protect\citeauthoryear{{Arzamasskiy} \& {Rafikov}}{{Arzamasskiy} \&
  {Rafikov}}{2017}]{2017arXiv171001304A}
{Arzamasskiy} L.,  {Rafikov} R.~R.,  2017, ArXiv e-prints, arXiv:1710.01304

\bibitem[\protect\citeauthoryear{{Bate}, {Lubow}, {Ogilvie} \& {Miller}}{{Bate}
  et~al.}{2003}]{2003MNRAS.341..213B}
{Bate} M.~R.,  {Lubow} S.~H.,  {Ogilvie} G.~I.,    {Miller} K.~A.,  2003,
  \mnras, 341, 213

\bibitem[\protect\citeauthoryear{{Bitsch}, {Crida}, {Libert} \&
  {Lega}}{{Bitsch} et~al.}{2013}]{2013A&A...555A.124B}
{Bitsch} B.,  {Crida} A.,  {Libert} A.-S.,    {Lega} E.,  2013, \aap, 555, A124

\bibitem[\protect\citeauthoryear{{Bitsch} \& {Kley}}{{Bitsch} \&
  {Kley}}{2011}]{2011A&A...530A..41B}
{Bitsch} B.,  {Kley} W.,  2011, \aap, 530, A41

\bibitem[\protect\citeauthoryear{{Burns}}{{Burns}}{1976}]{1976AmJPh..44..944B}
{Burns} J.~A.,  1976, American Journal of Physics, 44, 944

\bibitem[\protect\citeauthoryear{{Chiang} \& {Goldreich}}{{Chiang} \&
  {Goldreich}}{1997}]{1997ApJ...490..368C}
{Chiang} E.~I.,  {Goldreich} P.,  1997, \apj, 490, 368

\bibitem[\protect\citeauthoryear{{Cresswell}, {Dirksen}, {Kley} \&
  {Nelson}}{{Cresswell} et~al.}{2007}]{2007A&A...473..329C}
{Cresswell} P.,  {Dirksen} G.,  {Kley} W.,    {Nelson} R.~P.,  2007, \aap, 473,
  329

\bibitem[\protect\citeauthoryear{{D'Angelo} \& {Lubow}}{{D'Angelo} \&
  {Lubow}}{2008}]{2008ApJ...685..560D}
{D'Angelo} G.,  {Lubow} S.~H.,  2008, \apj, 685, 560

\bibitem[\protect\citeauthoryear{{D'Angelo} \& {Lubow}}{{D'Angelo} \&
  {Lubow}}{2010}]{2010ApJ...724..730D}
{D'Angelo} G.,  {Lubow} S.~H.,  2010, \apj, 724, 730

\bibitem[\protect\citeauthoryear{{de Val-Borro}, {Artymowicz}, {D'Angelo} \&
  {Peplinski}}{{de Val-Borro} et~al.}{2007}]{2007A&A...471.1043D}
{de Val-Borro} M.,  {Artymowicz} P.,  {D'Angelo} G.,    {Peplinski} A.,  2007,
  \aap, 471, 1043

\bibitem[\protect\citeauthoryear{{Demianski} \& {Ivanov}}{{Demianski} \&
  {Ivanov}}{1997}]{1997A&A...324..829D}
{Demianski} M.,  {Ivanov} P.~B.,  1997, \aap, 324, 829

\bibitem[\protect\citeauthoryear{{Dong}, {Rafikov}, {Stone} \&
  {Petrovich}}{{Dong} et~al.}{2011}]{2011ApJ...741...56D}
{Dong} R.,  {Rafikov} R.~R.,  {Stone} J.~M.,    {Petrovich} C.,  2011, \apj,
  741, 56

\bibitem[\protect\citeauthoryear{{Dong}, {Zhu}, {Rafikov} \& {Stone}}{{Dong}
  et~al.}{2015}]{2015ApJ...809L...5D}
{Dong} R.,  {Zhu} Z.,  {Rafikov} R.~R.,    {Stone} J.~M.,  2015, \apjl, 809, L5

\bibitem[\protect\citeauthoryear{{Dorschner}, {Begemann}, {Henning}, {Jaeger}
  \& {Mutschke}}{{Dorschner} et~al.}{1995}]{1995A&A...300..503D}
{Dorschner} J.,  {Begemann} B.,  {Henning} T.,  {Jaeger} C.,    {Mutschke} H.,
  1995, \aap, 300, 503

\bibitem[\protect\citeauthoryear{{Duffell} \& {MacFadyen}}{{Duffell} \&
  {MacFadyen}}{2012}]{2012ApJ...755....7D}
{Duffell} P.~C.,  {MacFadyen} A.~I.,  2012, \apj, 755, 7

\bibitem[\protect\citeauthoryear{{Fragner} \& {Nelson}}{{Fragner} \&
  {Nelson}}{2010}]{2010A&A...511A..77F}
{Fragner} M.~M.,  {Nelson} R.~P.,  2010, \aap, 511, A77

\bibitem[\protect\citeauthoryear{{Goldreich} \& {Tremaine}}{{Goldreich} \&
  {Tremaine}}{1979}]{1979ApJ...233..857G}
{Goldreich} P.,  {Tremaine} S.,  1979, \apj, 233, 857

\bibitem[\protect\citeauthoryear{{Goldreich} \& {Tremaine}}{{Goldreich} \&
  {Tremaine}}{1980}]{GT80}
{Goldreich} P.,  {Tremaine} S.,  1980, \apj, 241, 425

\bibitem[\protect\citeauthoryear{{Goodman} \& {Rafikov}}{{Goodman} \&
  {Rafikov}}{2001}]{2001ApJ...552..793G}
{Goodman} J.,  {Rafikov} R.~R.,  2001, \apj, 552, 793

\bibitem[\protect\citeauthoryear{{Ju}, {Stone} \& {Zhu}}{{Ju}
  et~al.}{2016}]{Ju}
{Ju} W.,  {Stone} J.~M.,    {Zhu} Z.,  2016, \apj, 823, 81

\bibitem[\protect\citeauthoryear{{Ju}, {Stone} \& {Zhu}}{{Ju}
  et~al.}{2017}]{Ju2017}
{Ju} W.,  {Stone} J.~M.,    {Zhu} Z.,  2017, \apj, 841, 29

\bibitem[\protect\citeauthoryear{{Juh{\'a}sz}, {Benisty}, {Pohl}, {Dullemond},
  {Dominik} \& {Paardekooper}}{{Juh{\'a}sz} et~al.}{2015}]{2015MNRAS.451.1147J}
{Juh{\'a}sz} A.,  {Benisty} M.,  {Pohl} A.,  {Dullemond} C.~P.,  {Dominik} C.,
    {Paardekooper} S.-J.,  2015, \mnras, 451, 1147

\bibitem[\protect\citeauthoryear{{Lubow} \& {Ogilvie}}{{Lubow} \&
  {Ogilvie}}{2000}]{2000ApJ...538..326L}
{Lubow} S.~H.,  {Ogilvie} G.~I.,  2000, \apj, 538, 326

\bibitem[\protect\citeauthoryear{{Masset} \& {Tagger}}{{Masset} \&
  {Tagger}}{1996}]{1996A&A...307...21M}
{Masset} F.,  {Tagger} M.,  1996, \aap, 307, 21

\bibitem[\protect\citeauthoryear{{Matsakos} \& {K{\"o}nigl}}{{Matsakos} \&
  {K{\"o}nigl}}{2017}]{2017AJ....153...60M}
{Matsakos} T.,  {K{\"o}nigl} A.,  2017, \aj, 153, 60

\bibitem[\protect\citeauthoryear{{Nelson}, {Gressel} \& {Umurhan}}{{Nelson}
  et~al.}{2013}]{2013MNRAS.435.2610N}
{Nelson} R.~P.,  {Gressel} O.,    {Umurhan} O.~M.,  2013, \mnras, 435, 2610

\bibitem[\protect\citeauthoryear{{Nelson} \& {Papaloizou}}{{Nelson} \&
  {Papaloizou}}{1999}]{1999MNRAS.309..929N}
{Nelson} R.~P.,  {Papaloizou} J.~C.~B.,  1999, \mnras, 309, 929

\bibitem[\protect\citeauthoryear{{Ostriker}}{{Ostriker}}{1999}]{1999ApJ...513..252O}
{Ostriker} E.~C.,  1999, \apj, 513, 252

\bibitem[\protect\citeauthoryear{{Owen} \& {Lai}}{{Owen} \&
  {Lai}}{2017}]{2017arXiv170309250O}
{Owen} J.~E.,  {Lai} D.,  2017, ArXiv e-prints

\bibitem[\protect\citeauthoryear{{Papaloizou} \& {Lin}}{{Papaloizou} \&
  {Lin}}{1995}]{1995ApJ...438..841P}
{Papaloizou} J.~C.~B.,  {Lin} D.~N.~C.,  1995, \apj, 438, 841

\bibitem[\protect\citeauthoryear{{Petrovich} \& {Rafikov}}{{Petrovich} \&
  {Rafikov}}{2012}]{2012ApJ...758...33P}
{Petrovich} C.,  {Rafikov} R.~R.,  2012, \apj, 758, 33

\bibitem[\protect\citeauthoryear{{Rafikov}}{{Rafikov}}{2002}]{2002ApJ...569..997R}
{Rafikov} R.~R.,  2002, \apj, 569, 997

\bibitem[\protect\citeauthoryear{{Rafikov}}{{Rafikov}}{2016}]{2016ApJ...831..122R}
{Rafikov} R.~R.,  2016, \apj, 831, 122

\bibitem[\protect\citeauthoryear{{Rafikov} \& {Petrovich}}{{Rafikov} \&
  {Petrovich}}{2012}]{2012ApJ...747...24R}
{Rafikov} R.~R.,  {Petrovich} C.,  2012, \apj, 747, 24

\bibitem[\protect\citeauthoryear{{Rein}}{{Rein}}{2012}]{2012MNRAS.422.3611R}
{Rein} H.,  2012, \mnras, 422, 3611

\bibitem[\protect\citeauthoryear{{Sotiriadis}, {Libert}, {Bitsch} \&
  {Crida}}{{Sotiriadis} et~al.}{2017}]{2017A&A...598A..70S}
{Sotiriadis} S.,  {Libert} A.-S.,  {Bitsch} B.,    {Crida} A.,  2017, \aap,
  598, A70

\bibitem[\protect\citeauthoryear{{Tanaka}, {Takeuchi} \& {Ward}}{{Tanaka}
  et~al.}{2002}]{2002ApJ...565.1257T}
{Tanaka} H.,  {Takeuchi} T.,    {Ward} W.~R.,  2002, \apj, 565, 1257

\bibitem[\protect\citeauthoryear{{Tanaka} \& {Ward}}{{Tanaka} \&
  {Ward}}{2004}]{2004ApJ...602..388T}
{Tanaka} H.,  {Ward} W.~R.,  2004, \apj, 602, 388

\bibitem[\protect\citeauthoryear{{Ward}}{{Ward}}{1986}]{1986Icar...67..164W}
{Ward} W.~R.,  1986, \icarus, 67, 164

\bibitem[\protect\citeauthoryear{{Xiang-Gruess} \& {Papaloizou}}{{Xiang-Gruess}
  \& {Papaloizou}}{2013}]{2013MNRAS.431.1320X}
{Xiang-Gruess} M.,  {Papaloizou} J.~C.~B.,  2013, \mnras, 431, 1320

\bibitem[\protect\citeauthoryear{{Zhu}, {Dong}, {Stone} \& {Rafikov}}{{Zhu}
  et~al.}{2015}]{2015ApJ...813...88Z}
{Zhu} Z.,  {Dong} R.,  {Stone} J.~M.,    {Rafikov} R.~R.,  2015, \apj, 813, 88

\end{thebibliography}

\bsp

\label{lastpage}
\end{document}